\documentclass[fleqn,usenatbib]{mnras}

\usepackage{newtxtext,newtxmath}
\usepackage[T1]{fontenc}

\DeclareRobustCommand{\VAN}[3]{#2}
\let\VANthebibliography\thebibliography
\def\thebibliography{\DeclareRobustCommand{\VAN}[3]{##3}\VANthebibliography}

\usepackage{graphicx}	% Including figure files
\usepackage{amsmath}	% Advanced maths commands
\usepackage{amssymb}	% Extra maths symbols

\def\lesssim{\mathrel{\hbox{\rlap{\hbox{\lower4pt\hbox{$\sim$}}}\hbox{$<$}}}}
\def\gtrsim{\mathrel{\hbox{\rlap{\hbox{\lower4pt\hbox{$\sim$}}}\hbox{$>$}}}}

\newcommand{\msun}{\,\mbox{M$_{\odot}$}}

\newcommand{\kms}{\mbox{\,$\rm{km}\,s^{-1}$}}

\newcommand{\feii}{\mbox{Fe\,{\sc ii}}}
\newcommand{\feiii}{\mbox{Fe\,{\sc iii}}}
\newcommand{\Ni}{\mbox{$^{56}$Ni}}
\newcommand{\Co}{\mbox{$^{56}$Co}}
\newcommand{\Fe}{\mbox{$^{56}$Fe}}

\newcommand{\NaI}{Na\,{\sc i}}
\newcommand{\CaII}{Ca\,{\sc ii}}

\newcommand{\MgII}{Mg\,{\sc ii}}
\newcommand{\SiII}{Si\,{\sc ii}}
\newcommand{\SiIII}{Si\,{\sc iii}}
\newcommand{\SrII}{Sr\,{\sc ii}}
\newcommand{\BaII}{Ba\,{\sc ii}}
\newcommand{\MoII}{Mo\,{\sc ii}}
\newcommand{\RuII}{Ru\,{\sc ii}}
\newcommand{\OII}{O\,{\sc ii}}
\newcommand{\OI}{O\,{\sc i}}
\newcommand{\CII}{C\,{\sc ii}}
\newcommand{\SII}{S\,{\sc ii}}

\newcommand{\ergs}{\,erg\,s$^{\mathrm{-1}}$}

\usepackage{xcolor}
\usepackage[utf8]{inputenc}
\usepackage[english]{babel}
\usepackage{multirow}
\usepackage{natbib}

\usepackage{booktabs,threeparttable}

\title[AT2018kzr: the merger of a WD and NS/BH]{AT2018kzr: the merger of an oxygen-neon white dwarf and a neutron star or black hole}

\author[J. H. Gillanders et al.]{
J. H. Gillanders,\thanks{E-mail: jgillanders01@qub.ac.uk}
S. A. Sim, 
S. J. Smartt
\\
Astrophysics Research Centre, School of Mathematics and Physics, Queen’s University Belfast, BT7 1NN, UK\\
}

\date{Accepted XXX. Received YYY; in original form ZZZ}

\pubyear{2020}

\begin{document}
\label{firstpage}
\pagerange{\pageref{firstpage}--\pageref{lastpage}}
\maketitle

% Abstract of the paper
\begin{abstract}
We present detailed spectroscopic analysis of the extraordinarily fast-evolving transient AT2018kzr. The transient's observed lightcurve showed a rapid decline rate, comparable to the kilonova AT2017gfo. We calculate a self-consistent sequence of radiative transfer models (using $\textsc{tardis}$) and determine that the ejecta material is dominated by intermediate-mass elements (O, Mg and Si), with a photospheric velocity of $\sim$\,12000--14500\kms. The early spectra have the unusual combination of being blue but dominated by strong \feii\ and \feiii\ absorption features. We show that this combination is only possible with a high Fe content (3.5\%). This implies a high Fe/(Ni+Co) ratio. Given the short time from the transient's proposed explosion epoch, the Fe cannot be \Fe\ resulting from the decay of radioactive \Ni\ synthesised in the explosion. Instead, we propose that this is stable $^{54}$Fe, and that the transient is unusually rich in this isotope. We further identify an additional, high-velocity component of ejecta material at $\sim$\,20000--26000\kms, which is mildly asymmetric and detectable through the \CaII\,NIR triplet. We discuss our findings with reference to a range of plausible progenitor systems and compare with published theoretical work. We conclude that AT2018kzr is most likely the result of a merger between an ONe white dwarf and a neutron star or black hole. As such, it would be the second plausible candidate with a good spectral sequence for the electromagnetic counterpart of a compact binary merger, after AT2017gfo.
\end{abstract}

% Select between one and six entries from the list of approved keywords.
% Don't make up new ones.

%%%%%%%%%%%%%%%%%%%%%%%%%%%%%%%%%%%%%%%%%%%%%%%%%%

%%%%%%%%%%%%%%%%% BODY OF PAPER %%%%%%%%%%%%%%%%%%

\section{Introduction} \label{Introduction}

\par
The advent of wide-field surveys and rapid multi-wavelength follow-up has produced discoveries of transients with much faster evolving lightcurves and spectra than the normal \Ni\ powered type Ia and Ibc population \citep{Poznanski2010,Drout2013,De2018,Prentice2018,Chen2020,Prentice2020}. The nature of these unusual transients is much debated and while many are explosions of a single star (albeit with a binary companion that influences the preceding evolution), some fraction may be violent stellar mergers. Extensive theoretical effort has been invested in exploring the mergers of compact objects, and binary combinations of white dwarfs (WDs), neutron stars (NSs) and stellar mass black holes (BHs) have all been computationally explored in various scenarios \citep{Fryer1999,King2007,Fernandez2019,Metzger2019}. Binary BH mergers are being discovered by LIGO-Virgo at a rate of one per week \citep{Abbott2016BBH1,AbbottGWTC1}, a candidate NS--BH merger has been discovered \citep[s190814bv;][]{GCN25324}, and detections of two binary NS mergers have been published \citep{LigoVirgo2017,Abbott2020a}.  Only one compact binary merger, which produced the signal GW170817, has been unambiguously linked to an electromagnetic signal \citep{AbbottMM2017}. Since WDs have radii that are $\sim$\,200--700 times larger than NS radii and Schwarzchild radii of 10\,\msun\,BHs, a binary merger with a WD component will radiate gravitational waves at frequencies below the LIGO-Virgo detectable range.

\par
There are no confirmed electromagnetic transients of WD--NS or WD--BH mergers to date. However, there is observational evidence for the progenitor binary systems existing. \cite{Miller-Jones2015} and \cite{Bahramian2017} present observational evidence for a candidate WD--BH binary system in the Galactic globular cluster 47 Tucanae. Theory predicts that these mergers are likely to be luminous \citep{Metzger2012,Margalit2016,Fernandez2019} and detectable in contemporary wide-field surveys. \cite{Zenati2020} predict somewhat fainter and redder optical transients than \cite{Metzger2012,Margalit2016,Fernandez2019}, but still comfortably within the sensitivity of ZTF \citep[][]{Bellm2019} and ATLAS \citep[][]{Smith2020} in the local Universe. The calculated rates of compact mergers with a WD component suggest they could be as high as 3--15\% of the type Ia rate \citep{Toonen2018}. With a {\em discovered} type Ia rate of $\sim$\,90 per year within 100\,Mpc \citep[][]{Smith2020}, such a high rate would imply there should be 2--13 such events per year, within this local volume.  Given that compact binary mergers have been detected through GW signals, and the large theoretical rate of mergers with a WD component, it is surprising that there are no obvious candidates from the wide-field surveys. Significant computational work has also been published on the merger of WDs with intermediate mass black holes \citep[][]{Rosswog2009,MacLeod2016} but no convincing candidate has yet been identified. 

\par
The most reliable method of determining the chemical composition of an astrophysical transient is by analysing its spectra with radiative transfer methods. From spectral modelling, one can aim to deduce the composition of the spectral forming region. This information is invaluable when it comes to determining the progenitor system, and understanding the physical nature of an explosive transient. The open source radiative transfer code $\textsc{tardis}$ \citep{Kerzendorf2014,Kerzendorf2019} is now applied to model spectral sequences of both type Ia and type II supernovae \citep[][]{Vogl2020}. For SN Ia, it has been used to determine the ejecta composition of type Iax SNe to aid understanding of the explosion mechanisms \citep{Magee2016,Barnabas2017}, and to constrain the existence of He in double detonation scenarios \citep{Boyle2017}.

\par
In this paper, we present a detailed spectroscopic analysis of AT2018kzr, a remarkably fast declining transient jointly discovered by the Zwicky Transient Facility and ATLAS, as described in \cite{McBrien2019}. At certain epochs in its lightcurve it fades almost as fast as the kilonova AT2017gfo, and is similar to, or more extreme than, the equally interesting SN2019bkc \citep[][]{Prentice2020,Chen2020}.  It was originally classified as a type Ic supernova on the Transient Name Server \citep{Pineda2018} and \cite{McBrien2019} used the SN2018kzr designation due to the possibility of it being a supernova type event. Here we propose to relabel it AT2018kzr as we show it is more likely to arise from a merger, than a supernova type event.

\par
We summarise the spectral data used in this paper, previously published by \cite{McBrien2019}, in Section \ref{Data}. We used the radiative transfer code $\textsc{tardis}$ \citep{Kerzendorf2014,Kerzendorf2019} to model the ejecta in the early spectra, and estimate the chemical composition of this material (see Section \ref{Photospheric phase spectra}). In Section \ref{Post-photospheric phase spectra}, we present the results of our analysis of the late-time spectra. Section \ref{Discussion and interpretive analysis} contains additional analysis of our $\textsc{tardis}$ models, and in Section \ref{Discussion of progenitor systems}, we use our findings to infer the likelihood of different progenitor systems producing this extraordinary transient, particularly within the context of compact binary mergers. We conclude in Section \ref{Conclusions}.

\section{Data} \label{Data}

\par
\begin{figure}
    \centering
    \includegraphics[width=\linewidth]{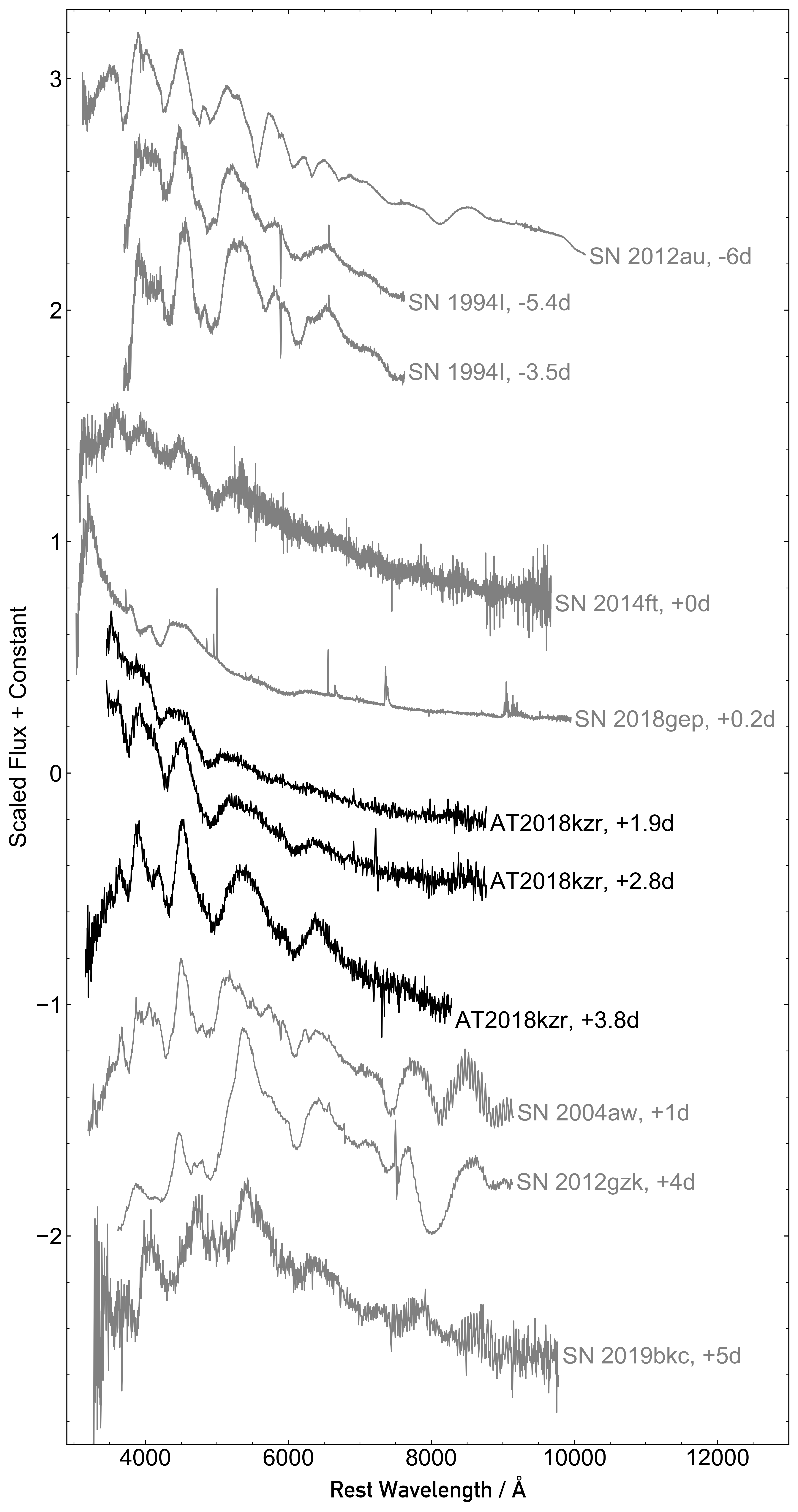}
    \caption{Photospheric spectra of AT2018kzr plotted alongside spectra of some fast-evolving type Ic SNe. The spectra have features in common with AT2018kzr, helping to corroborate our line identifications. However, it is clear that the spectral shape of these other objects are redder than the early spectra of AT2018kzr. The names and phases of each SN is listed next to the spectra. The phases are relative to V-band maximum \citep[SN1994I,][]{Yokoo1994}, B-band maximum \citep[SN2004aw,][]{Taubenberger2006}, R-band maximum \citep[SN2012au,][]{Takaki2013}, $r$-band maximum \citep[SN2012gzk, SN2014ft, SN2018gep,][]{Ben-Ami2012, De2018, Ho2019}, $g$-band maximum \citep[SN2019bkc,][]{Prentice2020}, and discovery (AT2018kzr).}
    \label{Figures/AT2018kzr_vs_Ic_SNe.png}
\end{figure}

The optical spectra used in this paper were presented in the discovery paper of \cite{McBrien2019}. We initially focused our effort on the three spectra of AT2018kzr obtained at phases of +1.895, +2.825 and +3.750 days after discovery. These were chosen as they were high signal-to-noise, reliably flux calibrated, and the transient remained in the photospheric phase. All three were taken at the ESO New Technology Telescope (NTT) as part of the extended Public ESO Spectroscopic Survey of Transient Objects: ePESSTO\footnote{www.pessto.org} \citep{Smartt2015}. We employed a refined version of the +3.750 day spectrum, reduced and calibrated as part of the upcoming ePESSTO Phase 3 public data release which will be available through the ESO Science Archive Facility later in 2020. This data product has better correction of the fringing induced noise beyond 7000\,\AA\ and we ensured it was flux calibrated to the photometry in \cite{McBrien2019}. We also performed some analysis on two of the later spectra, where the transient appeared to no longer be in the photospheric phase; specifically, the +7.056 and the +14.136 day spectra taken with LRIS on the Keck I telescope. The two LRIS Keck spectra were taken at epochs when the transient had faded to comparable (or lower) flux levels to the host galaxy, and so careful host subtraction was required to recover an uncontaminated transient object spectrum. The methodology for the host subtraction is detailed in Appendix \ref{Appendix - Host galaxy subtraction for the late epoch spectra}. In order to place the spectral characteristics of AT2018kzr in context, we plot the spectra of objects with similar line features in Figure \ref{Figures/AT2018kzr_vs_Ic_SNe.png}. The line transitions we observe appear similar to some type Ic SNe, but with important differences in terms of the temperature at which the features appear. We will discuss the comparisons further in the relevant analysis sections. All five observational spectra analysed are shown in Figure \ref{Figures/sequence_of_models_vs_observations.png}.

\section{Photospheric phase spectra} \label{Photospheric phase spectra}

\par
Our main aim was to produce a sequence of self-consistent models, with a uniform, one-zone composition, that consistently reproduce the first three spectra. To do this, we followed the abundance tomography technique \citep[first presented by][]{Stehle2005}. We used $\textsc{tardis}$ \citep{Kerzendorf2014, Kerzendorf2019}, which is a one-dimensional Monte-Carlo radiative transfer code for generating synthetic supernova spectra. The code assumes that supernovae have an opaque core, from which $r$-packets (analogous to bundles of photons) are radiated, with frequencies randomly assigned based on the characteristic blackbody temperature of an effective photosphere that is assumed to lie at the boundary of the opaque core. These $r$-packets propagate through the ejecta material, which is assumed to be in homologous expansion, and interact with the matter (either via scattering or absorption). The emergent packets are used to compute the synthetic spectrum.

\par
Given how rapidly the transient evolved, we were not confident that any spectra taken at epochs later than the three NTT spectra could be meaningfully modelled with the assumption of an optically thick region and photosphere. $\textsc{tardis}$ is only applicable when modelling supernovae spectra in the photospheric phase, hence our decision to focus only on the three NTT spectra. We analyse the later spectra with an alternative method in Section \ref{Post-photospheric phase spectra}. An initial attempt to model the +2.8\,d spectrum was made by \cite{McBrien2019} to determine, approximately, the bulk composition of the ejecta. They determined that a composition dominated by O, Mg and Si, along with a small amount of Fe, could reproduce the observed absorption features. In this paper we go significantly further and calculate a detailed and consistent series of models to fit all of the early spectra.

\par
There are three important observational parameters from the observed lightcurve of AT2018kzr \citep[as presented by][]{McBrien2019} that provide constraints for our spectral models. First, because the transient was so rapidly declining, the system had to have a low ejecta mass. \cite{McBrien2019} estimated an ejecta mass of $M_{\rm ej} = 0.10\pm0.05$\,\msun, with which our spectral models should be compatible. Second, the constraints on the explosion epoch and the lightcurve modelling fits predict an explosion epoch $\sim$\,1.7\,d before first detection. Third, the bolometric luminosity of the transient is measured from the combined $griz$ photometry.

\par
\begin{figure}
    \centering
    \includegraphics[width=\linewidth]{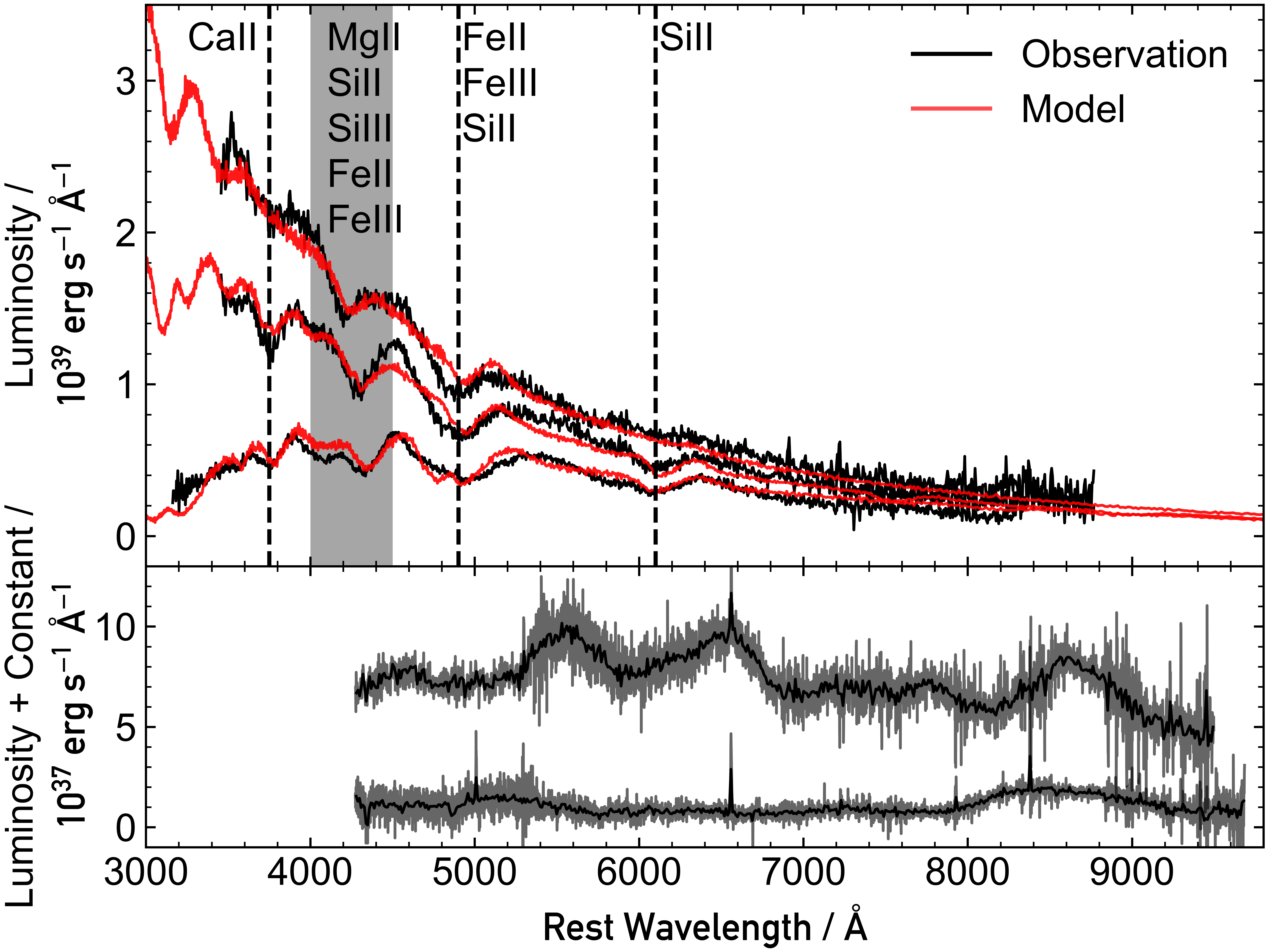}
    \caption{Observed spectra that were analysed in this work. \emph{Top Panel:} Comparison of the three photospheric phase spectra (earliest at the top and latest at the bottom), which have been fully reduced and flux-calibrated (black), with the corresponding models (red). Note that some regions of strong absorption by different ions have been annotated on the plot, to highlight what the dominant ion(s) are in our models at those wavelengths. \emph{Bottom Panel:} The two Keck spectra, taken at +7\,d (upper spectrum) and +14\,d (lower spectrum). These spectra have been rebinned to a dispersion of 10\,\AA\,pix$^{-1}$ (original in grey, and rebinned over-plotted in black).}
    \label{Figures/sequence_of_models_vs_observations.png}
\end{figure}

We adopted a power law density profile for our models, which has the general form:
\begin{equation}
\rho(v,t_{\rm exp}) = \rho_{\rm 0}\left(\frac{t_{\rm 0}}{t_{\rm exp}}\right)^{3}\left(\frac{v}{v_{\rm 0}}\right)^{-\Gamma}
\label{Density Profile Equation}
\end{equation}
for ${v_{\rm min}} < v < {v_{\rm max}}$, where $\rho_{\rm 0}$, $t_{\rm 0}$, $v_{\rm 0}$, $\Gamma$ and ${v_{\rm max}}$ are constants. The values for the constants in this equation were empirically derived as part of our efforts to fit the data, and are given in Table \ref{Table of Density Profile Parameters}. The only values in this equation that varied between our best fits for each epoch are minimum ejecta velocity, ${v_{\rm min}}$, and time since explosion, $t_{\rm exp}$. To illustrate the quality of the data and consistency of the final modelling results, we show the three best-fit models for the three spectra in Figure \ref{Figures/sequence_of_models_vs_observations.png}.

\begin{table*}
\centering
\caption{Velocity ranges and density profile parameters for each of the models at the three different epochs. The density profile parameters and outer velocity boundary used for each epoch remained the same, while the inner velocity boundary shifted for each epoch.}
\begin{tabular}{ccccccc}
\hline
\hline
\multirow{2}{*}{\begin{tabular}[c]{@{}c@{}}Time After \\ Discovery / Days\end{tabular}}   &\multirow{2}{*}{\begin{tabular}[c]{@{}c@{}}Minimum\\ Velocity / \kms\end{tabular}}   &\multirow{2}{*}{\begin{tabular}[c]{@{}c@{}}Maximum\\ Velocity / \kms\end{tabular}}  &\multicolumn{4}{c}{Density Profile Parameters}   \\
\cline{4-7}
&   &   &$\mathrm{v_{0}}$ / \kms    &$\mathrm{t_{0}}$ / Days    &$\mathrm{\rho_{0}}$ / g $\mathrm{cm^{-3}}$    &$\Gamma$ \\
\hline
+1.895   &13800   &25000   &14000    &2    &$10^{-12}$    &10     \\
+2.825   &12000   &25000   &14000    &2    &$10^{-12}$    &10     \\
+3.750   &14500   &25000   &14000    &2    &$10^{-12}$    &10     \\
\hline
\end{tabular}
\label{Table of Density Profile Parameters}
\end{table*}

\par
\cite{McBrien2019} predict the explosion epoch to be $\lesssim$\,3 days prior to first detection. However, we found that the spectra were fit best by using a model explosion epoch 4.6 days prior to first detection. We do not consider this result contradictory to that of \cite{McBrien2019}, as they only weakly constrained the explosion epoch from the photometric lightcurve.  The most recent non-detection was two days before first detection and was relatively shallow. Hence we could comfortably accommodate a slower rise time than that inferred by their bolometric lightcurve. Table \ref{Table of Model Properties} contains the time since explosion used for each of our best-fit models.

\begin{table*}
\centering
\caption{$\textsc{tardis}$ model parameters compared with those inferred from the photospheric lightcurve. Epoch refers to the estimated time from explosion from the lightcurve, whereas $\mathrm{t_{exp}}$ is that inferred from $\textsc{tardis}$. The $\mathrm{T_{eff}}$ is not a free parameter; it is calculated consistently in $\textsc{tardis}$ from the luminosity, velocity profile and $\mathrm{t_{exp}}$ in each model.}
\begin{tabular}{cccccc}
\hline
\hline
Epoch / Days   &{\begin{tabular}[c]{@{}c@{}}Model Time Since \\ Explosion, $\mathrm{t_{exp}}$ / Days\end{tabular}}&$\mathrm{T_{eff}}$ / K    &{\begin{tabular}[c]{@{}c@{}}Luminosity From \\ $\textsc{tardis}$ / \ergs \end{tabular}}     &{\begin{tabular}[c]{@{}c@{}}Luminosity From LC \\ Model / \ergs \end{tabular}}   &{\begin{tabular}[c]{@{}c@{}}Luminosity From \\ Photometry / \ergs \end{tabular}}   \\
\hline
+1.895    &+6.5    &13200    &1.30$\times10^{43}$   &1.44$\times10^{43}$    &N/A                             \\ 
+2.825    &+7.5    &12100    &9.22$\times10^{42}$   &5.99$\times10^{42}$    &(8.4$\pm$1.3)$\times10^{42}$    \\
+3.750    &+8.5    &8180     &3.62$\times10^{42}$   &2.64$\times10^{42}$    &(3.8$\pm$0.9)$\times10^{42}$    \\
\hline
\end{tabular}
\label{Table of Model Properties}
\end{table*}

\par
An important input parameter for $\textsc{tardis}$ is the luminosity of the transient at each epoch. For this, we used the bolometric lightcurve luminosities from \cite{McBrien2019}, at the epochs of the NTT spectra. We were able to obtain good fits to the observed spectra without deviating more than $\sim$\,50\% from these luminosity values.  The fact we have managed to fit the spectra with similar values to those presented by \cite{McBrien2019} further corroborates their luminosity estimates for AT2018kzr. Table \ref{Table of Model Properties} contains the luminosities used in each of our $\textsc{tardis}$ models. Also included are two different bolometric luminosities from \cite{McBrien2019}. The first are the values determined from the measured data points. The second are the values of a physical model fit to the data. Both sets of data are included to highlight the variation between the photometric points and the best-fit bolometric lightcurve, and to enable comparisons between the luminosity obtained from our $\textsc{tardis}$ models, and the observed luminosity values.

\par
In abundance tomography studies, it is expected that the inner ejecta velocity should either remain constant or decrease with time. This corresponds to the photosphere being either stationary or receding \citep[see][for some examples]{Stehle2005, Hachinger2009}. However in the case of AT2018kzr, our best-fit models have an inner ejecta velocity that decreases from the first to second epoch, but then increases out from the second to third epoch. At face value, this would imply that the photosphere has moved outward. This may be possible in certain cases where, due to a change in density and temperature as the model evolves, the optical depth of the ejecta drastically increases. However, it is more likely that the photospheric approximation we are using is becoming more unreliable with time, and is introducing this feature into our models. Therefore, we emphasise that we do not interpret this as a physical property of the transient, and would advise the reader against drawing any strong conclusions from this aspect of the models.

\par
Since the third photospheric spectrum was the most evolved, and contained the most features, we began by constraining the composition for this epoch first. We assumed a composition dominated by the most abundant intermediate-mass elements (IMEs: C, O, Mg, Si, Ca)  and iron-group elements (IGEs: Ti, Cr, Fe, Co, Ni). Once a good fit for the +3.8\,d spectrum had been obtained, the composition was kept constant, and the other parameters of the model were adapted to fit the earlier two spectra. Table \ref{Table of Composition} outlines the final composition we used to consistently fit the three NTT spectra. Also included are estimated uncertainties for the elemental mass fractions. These represent the upper and lower values that we could employ for the element without significantly altering the quality of the fit across any of our models (see Section \ref{Oxygen as the dominant element} for further details of the O analysis in particular). Typically the uncertainties came from one of the three spectra which set the strongest constraints on an element (see Section \ref{Limits on carbon, sodium and sulphur}). As shown in \cite{McBrien2019}, the transient declined rapidly and in some phases the decline rate was similar to that of the one known kilonova, AT2017gfo. Hence we also attempted to model our spectra with a composition made from $r$-process elements, that would be plausible for a binary NS merger, and we discuss this in Section\,\ref{Ruling out trans-Fe group and r-process elements}. 

\begin{table}
\centering
\caption{Elemental composition of our best-fit model for the +1.9\,d spectrum. The composition remains uniform for all epochs, apart from the effect radioactive \Ni\ decay has on the composition. Also included are the limits we obtained for the maximum amounts of C and S that our best-fit models could tolerate. These results help us place an upper limit on the amount of the ejecta material that could be composed of these elements, which in turn can help us constrain the progenitor system.}
\begin{tabular}{ccc}
\hline
\hline
&Element   &Relative Mass Fraction       \\
\hline
&C     &< 0.08                           \\
&O     &0.7$^{+0.07}_{-0.15}$            \\
&Mg    &0.13$^{+0.10}_{-0.04}$           \\
&Si    &0.09$^{+0.03}_{-0.02}$           \\
&S     &< 0.02                           \\
&Ca    &(6$^{+2}_{-1})\times10^{-4}$     \\
&Ti    &(8$\pm0.8)\times10^{-4}$         \\
&Cr    &(5$^{+4}_{-2})\times10^{-4}$     \\
&Ni    &(2.5$\pm1)\times10^{-3}$         \\
&Co    &(3.5$^{+3}_{-1})\times10^{-3}$   \\
&Fe    &(3.5$\pm0.4)\times10^{-2}$       \\
\hline
\end{tabular}
\label{Table of Composition}
\end{table}

\par
\begin{figure*}
    \centering
    \includegraphics[width=\linewidth]{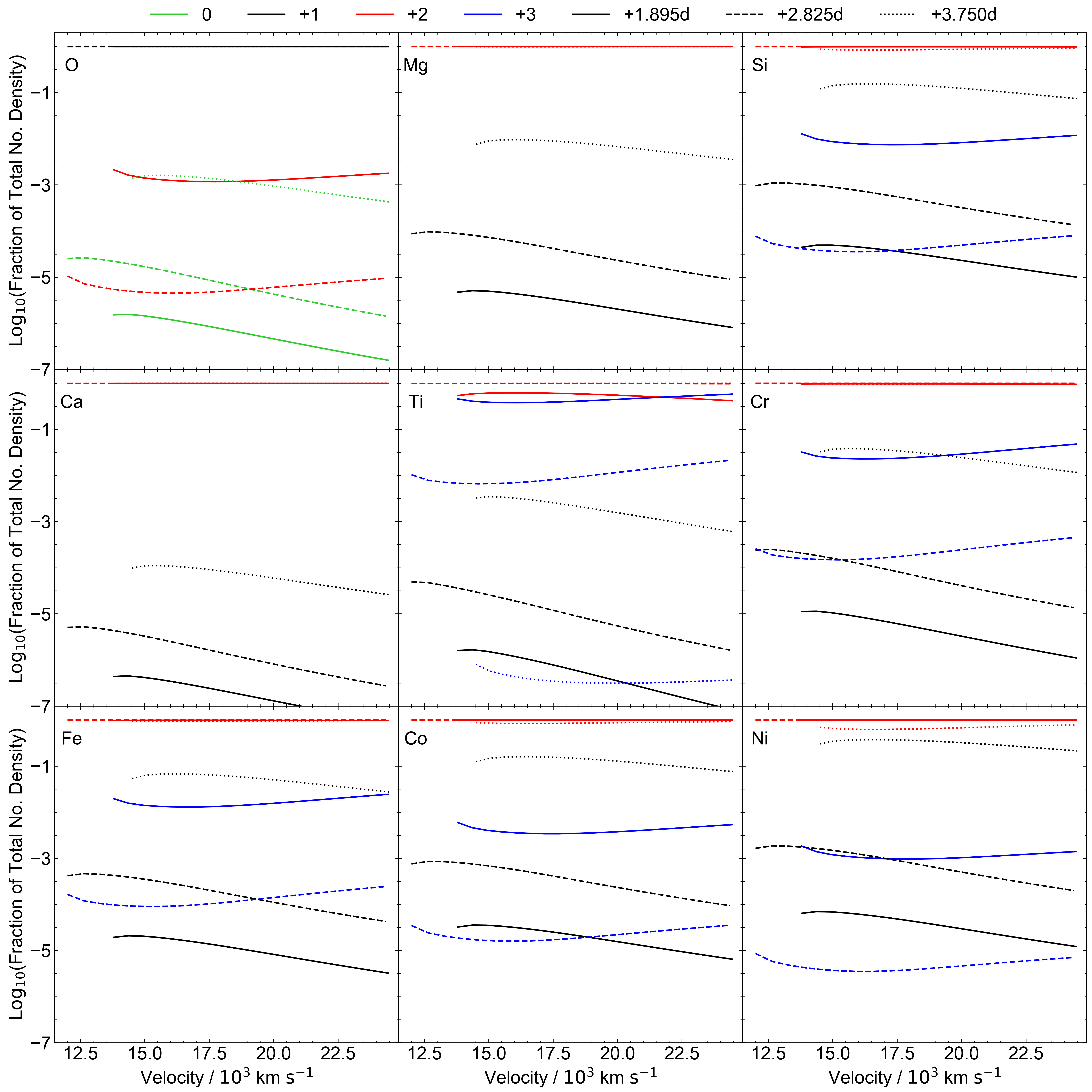}
    \caption{Ionisation states of all elements included in the models across the three epochs. The solid lines correspond to the first epoch, dashed to the second epoch, and dotted to the third. The lines are also colour-coded to indicate how ionised the ejected material is (neutral to triply ionised).}
    \label{Figures/multipanelled_ionisation_plot_new.pdf}
\end{figure*}

In Figure\,\ref{Figures/multipanelled_ionisation_plot_new.pdf}, we plot the number fraction of the different ions for each of the elements included in the model. This illustrates how the number fraction of each ion varies in the ejecta material, for each of the three photospheric epochs. This comparison helps to explain the spectral evolution of the models. As the transient evolves, it cools, and the elements present in the model become less ionised, resulting in lower ionisation species appearing and/or becoming more pronounced in the later spectra.

\subsection{Interpreting the +1.9 day spectrum} \label{Models for the +1.9 day spectrum}

\par
This spectrum has features in common with most of the type Ic SN spectra plotted in Figure \ref{Figures/AT2018kzr_vs_Ic_SNe.png}, which suggests a composition broadly similar to that of type Ic SNe, and this corroborates our line identifications. The SED of this spectrum is much bluer than most of these spectra (apart from that of SN2018gep), suggesting that AT2018kzr is hotter than the other transients at the epochs plotted. The two prominent absorption features are more pronounced in this spectrum than the equivalent features in the spectra of SN2018gep, SN2014ft and SN2019bkc. These spectral differences hint that AT2018kzr is fundamentally different from these other transients, which we confirm from preliminary modelling of their spectra (for example, we were able to satisfactorily fit the spectra of SN2014ft with a relative mass fraction of Fe ten times lower than that required for AT2018kzr).

\par
\begin{figure*}
    \centering
    \includegraphics[width=\linewidth]{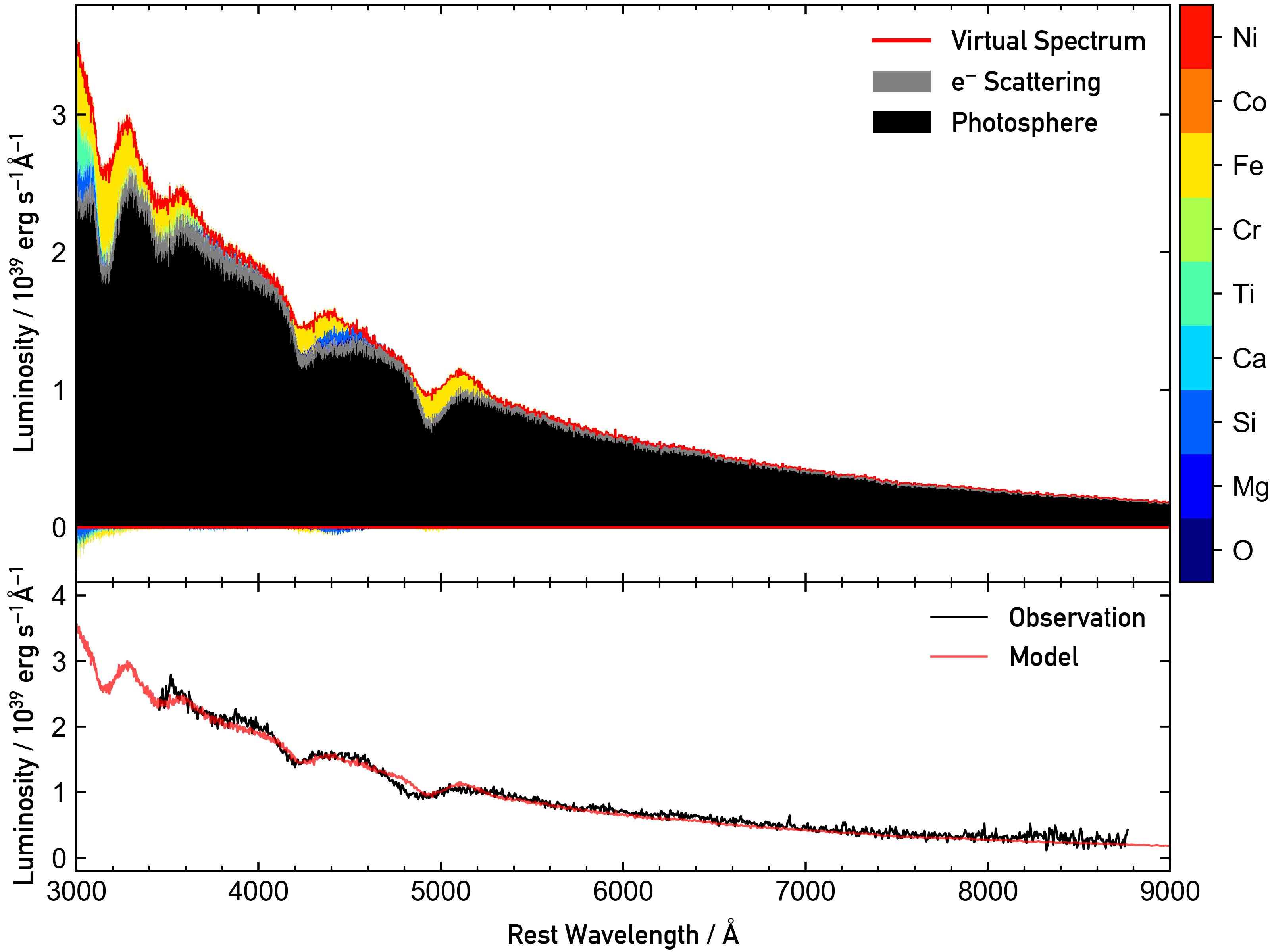}
    \caption{Best-fit model obtained for the +1.9\,d spectrum obtained for AT2018kzr. \emph{Bottom Panel:} Comparison of our best-fit model to the observed spectrum. There is good agreement between the observed spectrum and our model, with all the absorption and emission features broadly agreeing. \emph{Top Panel:} Associated Kromer plot for our model. The black region represents the contribution from the photosphere, and the grey represents the contribution from free electron scattering. The coloured regions correspond to the last physical interaction the escaping packets had; either absorption (region below zero) or emission (region above zero). This is a useful way to display which elements are contributing most to the various spectral features. At this epoch, the spectrum is dominated by Fe features (specifically \feiii).}
    \label{Figures/Epoch_1_spectral_comparison+kromer_plot.png}
\end{figure*}

The bottom panel of Figure \ref{Figures/Epoch_1_spectral_comparison+kromer_plot.png} illustrates our best-fit model plotted alongside the +1.9\,d spectrum.  The two prominent features in the observed spectrum are centred at $\sim$\,4200\,\AA\ and $\sim$\,4900\,\AA. Our model reproduces both of these features well. The upper panel of Figure \ref{Figures/Epoch_1_spectral_comparison+kromer_plot.png} highlights which elements contribute most to the absorption and emission features observed in the model spectrum. We define these plots as ``Kromer'' plots \citep[see Figure 6 in][for more information]{Kromer2012}. From our Kromer plot, we can see that Fe is contributing the most to both of these features. The feature at $\sim$\,4200\,\AA\ is dominated by the \feiii\ $\lambda \lambda$4396, 4420, 4431 lines, with contribution from the \SiIII\ $\lambda$4553 line. The feature at $\sim$4900 \AA\ is dominated by the \feiii\ $\lambda \lambda$5127, 5156 lines. See Table \ref{Table of Composition} for the full composition used in the models. 

\par
Beyond these two features, the rest of the spectrum is featureless. The important parameters of this $\textsc{tardis}$ model are given in Tables \ref{Table of Density Profile Parameters} and \ref{Table of Model Properties}, including time since explosion, effective photospheric temperature, and the inner luminosity. An important initial result from this spectrum is that it is unusually blue for the presence of strong \feiii\ lines. The combination of the high temperature and strong \feiii\ lines then requires a surprisingly high Fe abundance (see Table \ref{Table of Composition}).

\subsection{Interpreting the +2.8 day spectrum} \label{Models for the +2.8 day spectrum}

\par
This spectrum is almost as hot as the first epoch spectrum, and is still bluer than most of the Ic spectra in Figure \ref{Figures/AT2018kzr_vs_Ic_SNe.png} (apart from SN2018gep; comparable SED and hotter than SN2014ft). The absorption features are also more pronounced in this spectrum than the corresponding features in the other hot, blue spectra. Our model implies these are now blends of \feii\ and \feiii\ absorption, with contribution from \MgII. \feii\ has previously been invoked to explain these broad absorption lines in type Ic SNe \citep[e.g.][]{Hachinger2009}. However, our spectrum is clearly much bluer and hotter than all the comparison Ic spectra in Figure \ref{Figures/AT2018kzr_vs_Ic_SNe.png}, and it is unusual, if not unprecedented, to observe such a hot continuum with strong \feii\ and \feiii\ lines. From a purely empirical perspective, this suggests a larger Fe abundance, or higher density material (since a higher density equates to lower ionisation states, and more Fe atoms), would be required to fit these lines in our models. We note that the possibility of high typical density in AT2018kzr at peak is potentially broadly consistent with the fast rise time of AT2018kzr compared to normal Ic SNe. Specifically, the fast rise is suggestive of a low ejecta mass (rise time to peak brightness, $t_{\rm pk} \propto \sqrt{M_{\rm ej}}$), and, since the density of ejecta material at peak, $\rho_{\rm pk} \propto \frac{M_{\rm ej}}{t_{\rm pk}^3} \propto \frac{1}{t_{\rm pk}}$, all other properties remaining equal (expansion velocity, opacity, etc.), the characteristic density of AT2018kzr would be higher than that of normal Ic ejecta material.

\par
\begin{figure*}
    \centering
    \includegraphics[width=\linewidth]{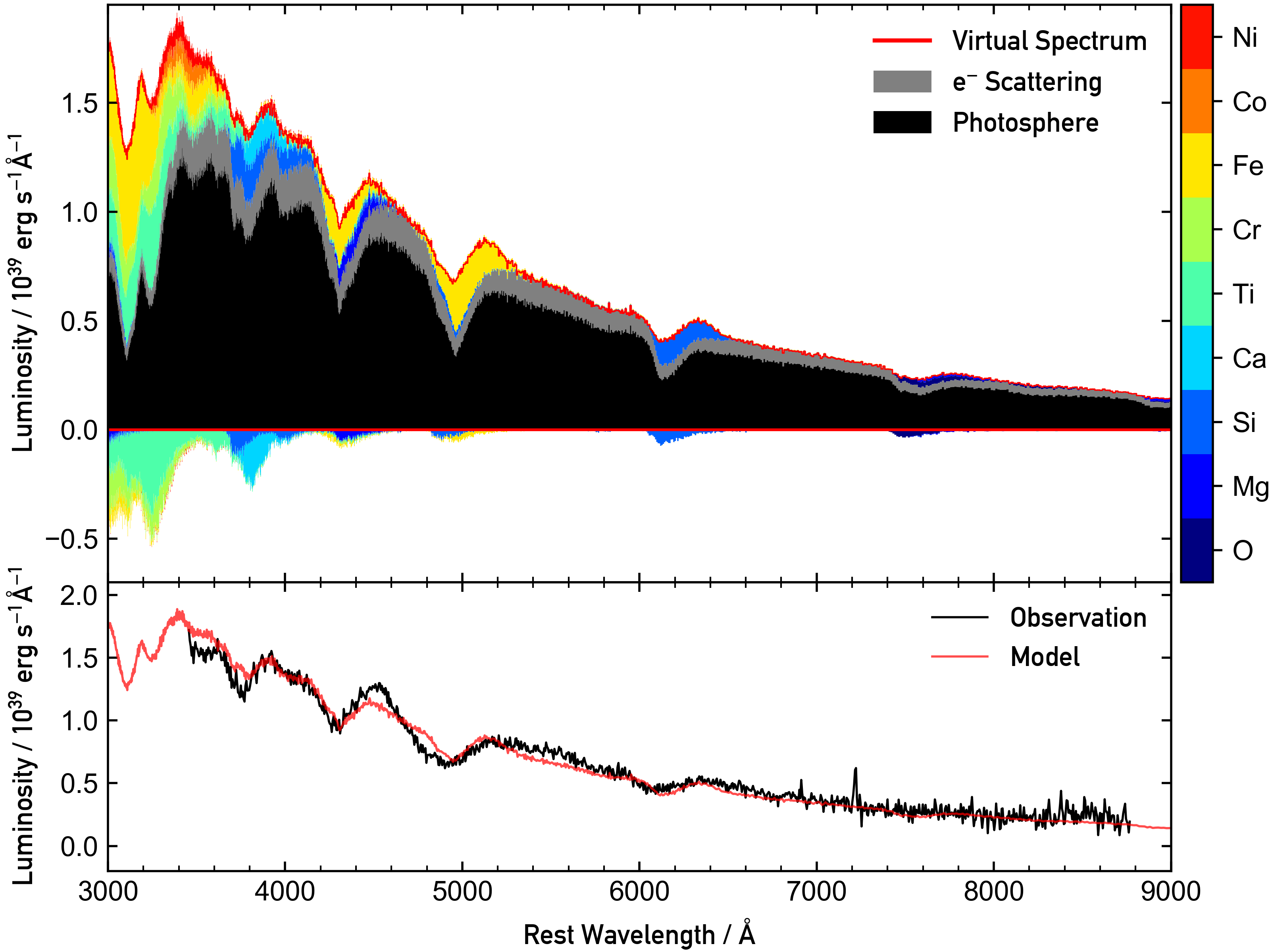}
    \caption{Best-fit model obtained for the +2.8\,d spectrum. \emph{Bottom Panel:} There is good agreement between the observed spectrum and our model, with all the absorption and emission features broadly agreeing, with the exception of the emission feature at $\sim$\,4600\,\AA\ being under-pronounced in our model. \emph{Top Panel:} Associated Kromer plot for our model. The absorption and emission features due to different elements can be seen. At this epoch, the spectrum is again dominated by Fe, but with more pronounced features from the other elements included in the model, as compared with the model in Figure \ref{Figures/Epoch_1_spectral_comparison+kromer_plot.png}.}
    \label{Figures/Epoch_2_spectral_comparison+kromer_plot.png}
\end{figure*}

The bottom panel of Figure \ref{Figures/Epoch_2_spectral_comparison+kromer_plot.png} shows the best-fit model plotted alongside the +2.8\,d spectrum. The two features present in the previous spectrum (centred on $\sim$\,4200\,\AA\ and $\sim$\,4900\,\AA) are still prominent at this epoch. Two additional features have also appeared, one centred at $\sim$\,3700\,\AA, and the other centred at $\sim$\,6100\,\AA. We were able to successfully reproduce these two new features with our model, and we identify these features as \CaII\ and \SiII, respectively.

\par
The two absorption features that were present in the first spectrum are still reproduced by the $\textsc{tardis}$ model. However, the features are no longer dominated by \feiii; the feature at $\sim$\,4200\,\AA\ is produced by significant contributions from the \MgII\ $\lambda \lambda$4481.1, 4481.3 lines, \SiII\ $\lambda$4131 line, and \feiii\ $\lambda \lambda$4420, 4431 lines, while the feature at $\sim$\,4900\,\AA\ is produced by the \SiII\ $\lambda \lambda$5041, 5056 lines, \feii\ $\lambda \lambda$4924, 5018, 5169 lines, and \feiii\ $\lambda \lambda$5127, 5156 lines. The ejecta have cooled sufficiently between the first and second epochs to allow the \feii/\feiii\ ratio to increase such that \feii\ now has a noticeable influence on the spectrum.

\subsection{Interpreting the +3.8 day spectrum} \label{Models for the +3.8 day spectrum}

\par
This spectrum looks very similar to some of the Ic spectra in Figure \ref{Figures/AT2018kzr_vs_Ic_SNe.png}. At this epoch, AT2018kzr has cooled and become noticeably redder, matching the SED of the other spectra. Without the previous two spectra, the rapid spectral evolution of AT2018kzr would not be evident, and this spectrum could be mistaken as belonging to a fast type Ic SN. However, these comparison objects do not exhibit the extremely rapid evolution we have observed for AT2018kzr. This hints at an alternative progenitor system that ejects material with a similar composition to that expected for type Ic SNe.

\par
\begin{figure*}
    \centering
    \includegraphics[width=\linewidth]{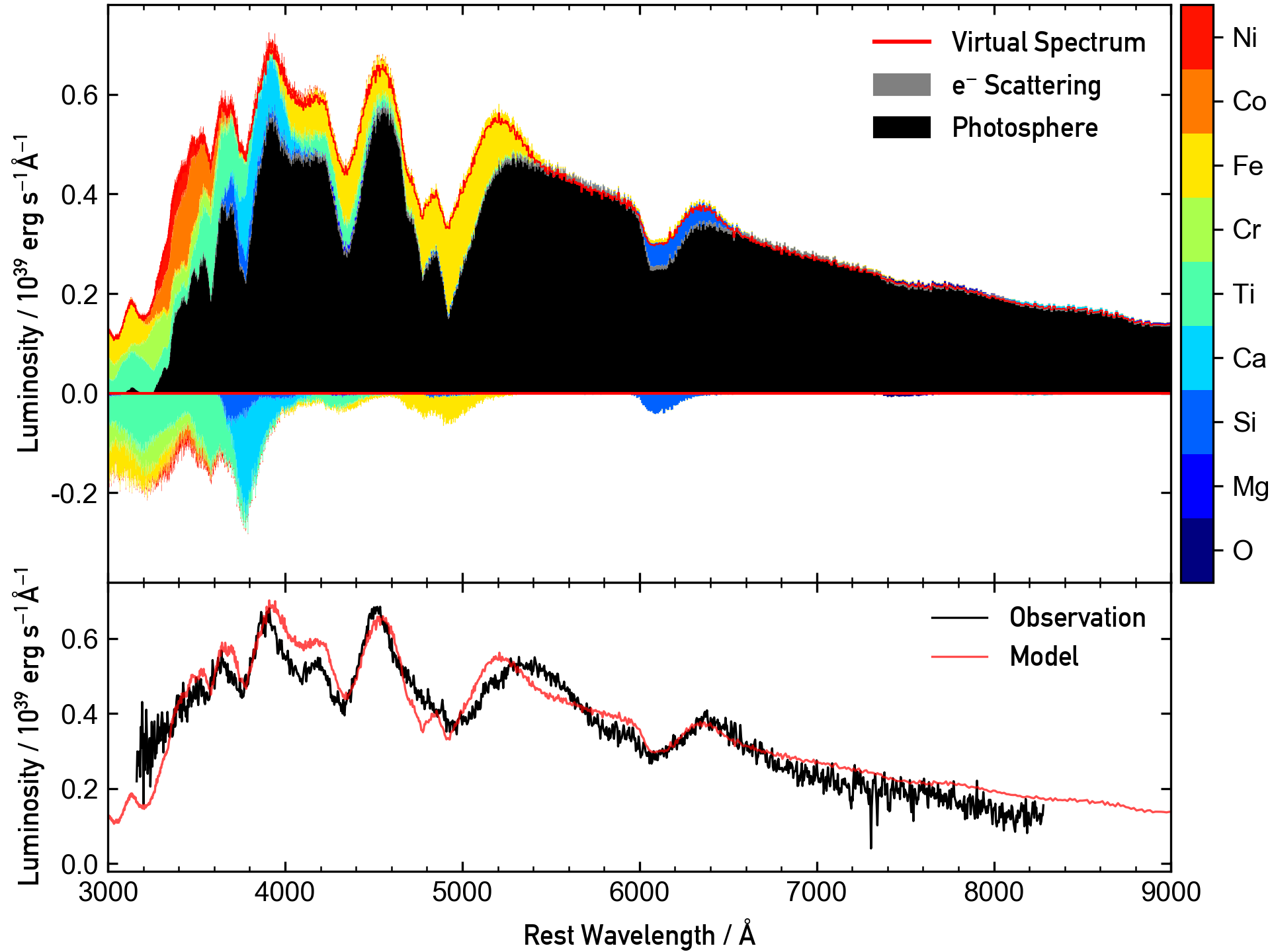}
    \caption{Best-fit model obtained for the +3.8\,d spectrum obtained for AT2018kzr. \emph{Bottom panel:} There is overall good agreement between the observed spectrum and our model, with all the absorption and emission features broadly agreeing. \emph{Top Panel:} Associated Kromer plot for our model. The absorption and emission features due to the different elements can be seen. At this epoch, the spectrum has contributions from multiple elements at the same wavelength, leading to blended features. There is clearly significant contributions from all of the elements included in the model, especially at the blue end.}
    \label{Figures/Epoch_3_spectral_comparison+kromer_plot.png}
\end{figure*}

The bottom panel of Figure \ref{Figures/Epoch_3_spectral_comparison+kromer_plot.png} compares this best-fit model with the +3.8\,d spectrum.  The features in this spectrum are more prominent than in the previous two epochs, and are again satisfactorily reproduced by our model. The feature at $\sim$\,3800\,\AA\ is well reproduced by \CaII, as before. The feature at $\sim$\,4200\,\AA\ is quantitatively reproduced by a blend of the \feii\ $\lambda \lambda$4233, 4523, 4549, 4584 lines (similar to the previous epoch, but with negligible contribution from \feiii, and less contribution from \MgII). The feature at $\sim$\,4900\,\AA\ is produced by the \feii\ $\lambda \lambda$4924, 5018, 5169, 5316 lines. Again, this is similar to the previous epoch, apart from the \feiii. Finally, the feature at $\sim$\,6100\,\AA\ is modelled adequately by purely \SiII, as before. The model has now expanded and cooled sufficiently for the contributions of \feiii\ to become negligible relative to \feii. In addition to the specific features we identify, there is significant line blanketing in the blue end of the observed spectrum below $\sim$\,3800\,\AA. In our model, we can replicate the strong absorption and reproduce the observed  blanketing with a mix of IGEs (Ti, Cr, Ni, Co and Fe).

\par
Our model slightly overestimates the flux towards the red end of the spectrum. Also, the absorption feature in the model at $\sim$\,4900\,\AA\ exhibits a double trough feature that is not present in the observed spectrum. There seems to be some discrepancy between our model and the observed spectrum in this region. Perhaps our model is missing an additional element/ion/transition that would blend these features. Alternatively, a change in velocity profile may blend these features into one. This may hint towards our model needing to have higher ejecta velocities, or a shallower density profile, allowing more material at higher ejecta velocities. Apart from this double trough feature, there is good agreement between the model and the observed spectrum, and so we do not place too much emphasis on this minor discrepancy.

\subsection{Modelling the photospheric spectra with an $r$-process element composition} \label{Ruling out trans-Fe group and r-process elements}

\par
Given the fast declining nature of the lightcurve and rapid spectral evolution we considered the possibility of AT2018kzr being a kilonova. \cite{McBrien2019} showed the lightcurve declined with a rate similar to AT2017gfo \citep[with the lightcurve compiled from data from][]{Andreoni2017, Arcavi2017, Chornock2017, Coulter2017, Cowperthwaite2017, Drout2017, Evans2017, Kasliwal2017, Smartt2017, Tanvir2017, Utsumi2017, Valenti2017}. We note that although AT2018kzr was significantly more luminous, it has been hypothesised that additional luminosity can be produced in a NS--NS merger from a newly formed massive NS \citep[see][]{Metzger2018}. Therefore, the lightcurve luminosity is not a sufficiently strong argument to rule out a kilonova. On this basis, we explored modelling the photospheric spectra with an alternative composition mixture that may be expected in binary NS mergers. 

\par
Kilonovae are expected to produce trans Fe-group and $r$-process elements, and so we generated models with the composition dominated by $r$-process elements from the first and second peaks (Kr--Ru and Te--Nd), to determine if the spectra of AT2018kzr exhibited any features due to these elements. We used the same explosion epoch as our previous models, presented in Table \ref{Table of Model Properties}, while we gave ourselves freedom to vary the luminosity and ejecta velocity. The relative abundances used in the models were based on their relative mass fractions as observed in the photosphere of the Sun \citep[from Table 1 in][]{Asplund2009} and we employed the atomic data from \cite{Kurucz1995} for these elements. It is critical to note that the atomic line lists are incomplete for these heavy elements, and so our models will not truly represent the entire range of features that these elements would produce, but we follow this strategy as an exploratory exercise.

\par
\begin{figure}
    \centering
    \includegraphics[width=\linewidth]{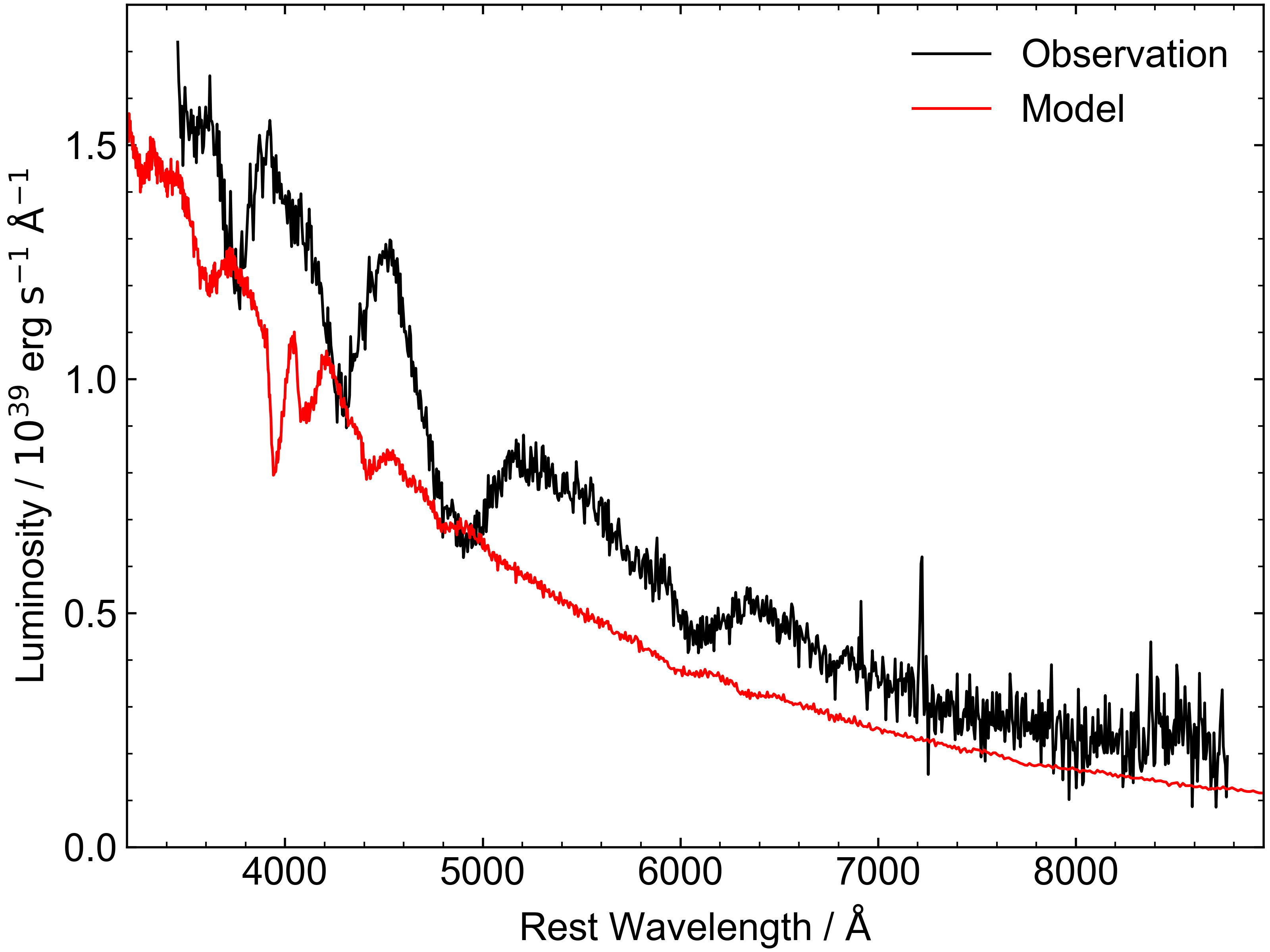}
    \caption{Comparison of one of our $r$-process model fits with the +2.8\,d spectrum. None of the observed features are replicated by the model, and there are features in the model that are not present in the observed spectrum.}
    \label{Figures/test_models_with_SoS_r-process_virtual_vs_observed_spectrum_zoomed.png}
\end{figure}

We were unable to produce a convincing fit with this composition and the parameters presented for our best-fit models, for any of the three photospheric spectra. The model which came closest to reproducing the general spectral shape in any epoch was for the +2.8\,d spectrum, and is shown in Figure \ref{Figures/test_models_with_SoS_r-process_virtual_vs_observed_spectrum_zoomed.png}. Due to the incompleteness of the line lists, we cannot definitively state that the model's inability to replicate observed features is proof that such a composition is ruled out. However, we can more convincingly determine that an $r$-process dominated composition is implausible based on the features produced in our models which are clearly not in the observed spectrum (particularly the features around $\sim$\,4000\,\AA, which are dominated by \SrII, \MoII, \RuII\ and \BaII).

\section{Post-photospheric phase spectra} \label{Post-photospheric phase spectra}

\par
In this section, we outline our analysis of the two Keck spectra (+7 and +14 days) shown in the bottom panel of Figure \ref{Figures/sequence_of_models_vs_observations.png}. Since these appear to be either transitioning to, or are within, the nebular phase, the use of $\textsc{tardis}$ for physical analysis is not justified. Therefore we employed simple, alternative methods of analysis. The +7\,d spectrum only has three discernible features, two of which have disappeared by the time the +14\,d spectrum was obtained. The feature at $\sim$\,7800--9000\,\AA\ appears to have a P-Cygni profile in the +7\,d spectrum. If this is the case, then there is still scattering of radiation, and the material is optically thick for the wavelength of this line. By the time the +14\,d spectrum was obtained, all trace of P-Cygni absorption has disappeared, and is replaced with a pure emission feature. This implies that the region which created the P-Cygni feature has become diffuse over time, and so the opacity for this line has dropped, preventing scattering. Since the two features cover the same wavelength range, we assumed they were produced by the same species, which we identified as the \CaII\ triplet. The \CaII\ triplet is commonly detected at late times in the spectra of many transients, and since the level that produces it has a low excitation energy, it can be easily excited. Additionally, it is excited from the ground state through the Ca H\&K resonance lines, and so it is commonly observed with even a small Ca mass and composition. There are no other plausible transitions at this wavelength to produce this feature.

\subsection{Interpreting the +7 day spectrum} \label{Interpreting the +7 day spectrum}

\par
We modelled the P-Cygni profile of the \CaII\ triplet using the generalised P-Cygni calculation code developed by U. Noebauer.\footnote{https://github.com/unoebauer/public-astro-tools} The code is based on a homologously expanding ejecta, as developed in the Elementary Supernova model \citep[as detailed by][]{Jeffery1990}. The program synthesises a model P-Cygni profile based on parameters specified by the user, which are the time since explosion, $t$, the optical depth of the line, $\tau$, the rest wavelength of the line, $\lambda_{0}$, two scaling velocities to infer the radial dependence of $\tau$, $v_{\rm e}$ and $v_{\rm ref}$, and the photospheric and maximum ejecta velocities, $v_{\rm phot}$ and $v_{\rm max}$, respectively. The photospheric velocity, or minimum ejecta velocity, is used to set the radius of the photosphere, which effectively sets the inner boundary for the model. In the context of the analysis we present here, we do not suppose to interpret this inner boundary as a true photosphere (which we would not expect to see at nebular phases). Instead we consider it as an approximate indication of the inner velocity boundary of the line forming region for the \CaII\ profile. Likewise, $v_{\rm max}$ provides a maximum velocity estimate for the line forming region.

\par
\begin{figure}
    \centering
    \includegraphics[width=\linewidth]{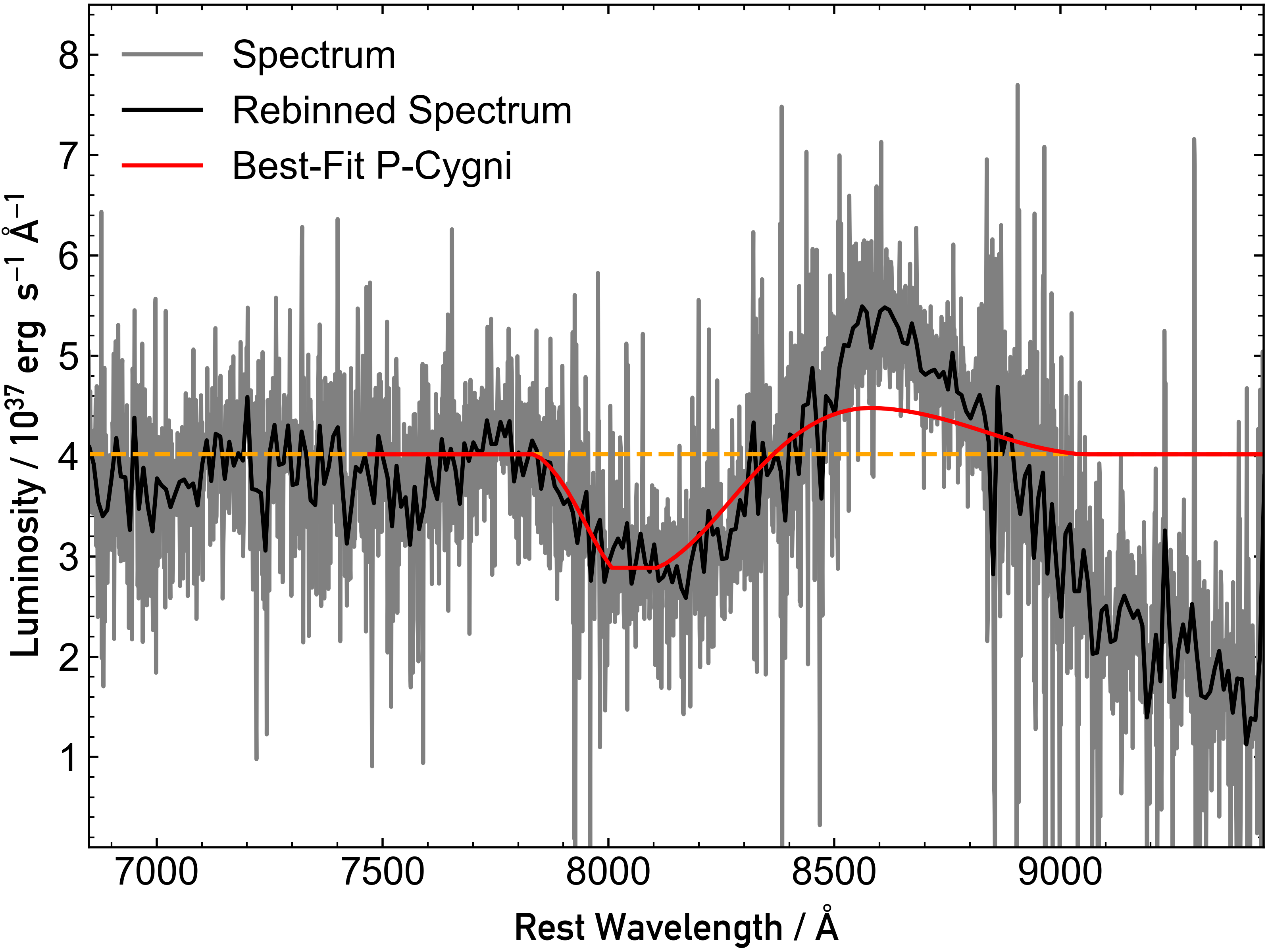}
    \caption{Model P-Cygni profile compared with the observed +7\,d spectrum. The orange dashed line marks the continuum.}
    \label{Figures/pcygni_plot_zoomed.png}
\end{figure}

Since several of these parameters are unknown and need to be treated as free variables, we amended the code to make a parameterised search for the best model P-Cygni profile to fit the data.
The values were varied over a plausible range, and the best fit was determined based on $\chi^{2}$ statistics.
The most physically interesting parameters are the minimum and maximum ejecta velocities, as well as the Sobolev optical depth, as these could be directly compared with those obtained from our $\textsc{tardis}$ models, enabling us to better understand the evolution of the transient. Table \ref{Table of post-photospheric spectra velocities} contains the parameters we obtained from our best-fit P-Cygni profile, which is shown in Figure \ref{Figures/pcygni_plot_zoomed.png}.

\begin{table*}
\centering
\caption{Best-fit model parameters for our analysis of the two late-time Keck spectra.}
\begin{tabular}{cccc}
\hline
\hline
Epoch / Days &{\begin{tabular}[c]{@{}c@{}}Minimum Ejecta\\Velocity, $v_{\rm phot}$ / \kms \end{tabular}} &{\begin{tabular}[c]{@{}c@{}}Maximum Ejecta\\Velocity, $v_{\rm max}$ / \kms \end{tabular}} &{\begin{tabular}[c]{@{}c@{}}Offset Velocity, \\ $v_{\rm offset}$ / \kms \end{tabular}}    \\
\hline
+7.056   &20000  &26000  &N/A    \\
+14.136  &N/A    &23700  &-2300  \\
\hline
\end{tabular}
\label{Table of post-photospheric spectra velocities}
\end{table*}

\par
To obtain velocity estimates, we need to fit the absorption trough. In Figure \ref{Figures/pcygni_plot_zoomed.png} it is clear that the P-Cygni model profile matches the observed spectrum in this region. The model profile is unable to reproduce the emission region of the observed spectrum. This is a result of the feature not being produced purely by scattering; clearly the feature is in net emission, and as a result we underestimate the amount of flux in this region. However, this is not important, as we are only interested in obtaining velocity ranges for the line-forming region.

\par
We obtained a minimum ejecta velocity, $v_{\rm phot}\sim$\,20000\kms, and a maximum ejecta velocity, $v_{\rm max}\sim$\,26000\kms. These velocity values are, surprisingly, $\sim$\,2 times higher than those obtained from our $\textsc{tardis}$ models. This is quite strong evidence that there is an additional high velocity component of the ejecta which is not observed in the early, photospheric spectra. The computation for the P-Cygni \CaII\ line produces a velocity from the absorption trough that is fairly robust, as is the velocity estimates from our $\textsc{tardis}$ models. Therefore we believe the difference to be a real physical effect, and is an indication of additional high velocity material, likely of low mass, which has been ejected beyond the emitting photospheric region (which is travelling at $\sim$\,12000--14500\kms).

\par
We note that the level of the observed continuum to the red of the P-Cygni peak falls significantly below that on the blue side. We attribute this to some over-subtraction of the host galaxy flux model as described in Appendix \ref{Appendix - Host galaxy subtraction for the late epoch spectra}, and do not attach any line forming significance to the discrepancy.

\par
Our best-fit model required a Sobolev optical depth, $\tau\sim$\,1.3, which tells us that the P-Cygni feature originates from a moderately optically thick region. Even though the +7\,d spectrum appears to be nebular, the region forming the P-Cygni feature is optically thick, at least for the wavelengths of the \CaII\ triplet. The optical depth for these lines in all three of our $\textsc{tardis}$ models are <\,1. Hence, the envelope is optically thin at these wavelengths in our $\textsc{tardis}$ models. However, the optical depth in the $\textsc{tardis}$ models becomes progressively larger with time. Since a P-Cygni line requires an optically thick transition, we have a consistent picture of the \CaII\ triplet line being optically thin at early epochs (probably due to high temperature), after which the material develops an optical depth around $\tau\sim$\,1.3, providing the opacity for the absorption trough to form. We will see the next spectrum is consistent with the idea that the opacity in this line has dropped further, due to expansion, and a pure emission component has formed.

\subsection{Interpreting the +14 day spectrum} \label{Interpreting the +14 day spectrum}

\par
\begin{figure}
    \centering
    \includegraphics[width=\linewidth]{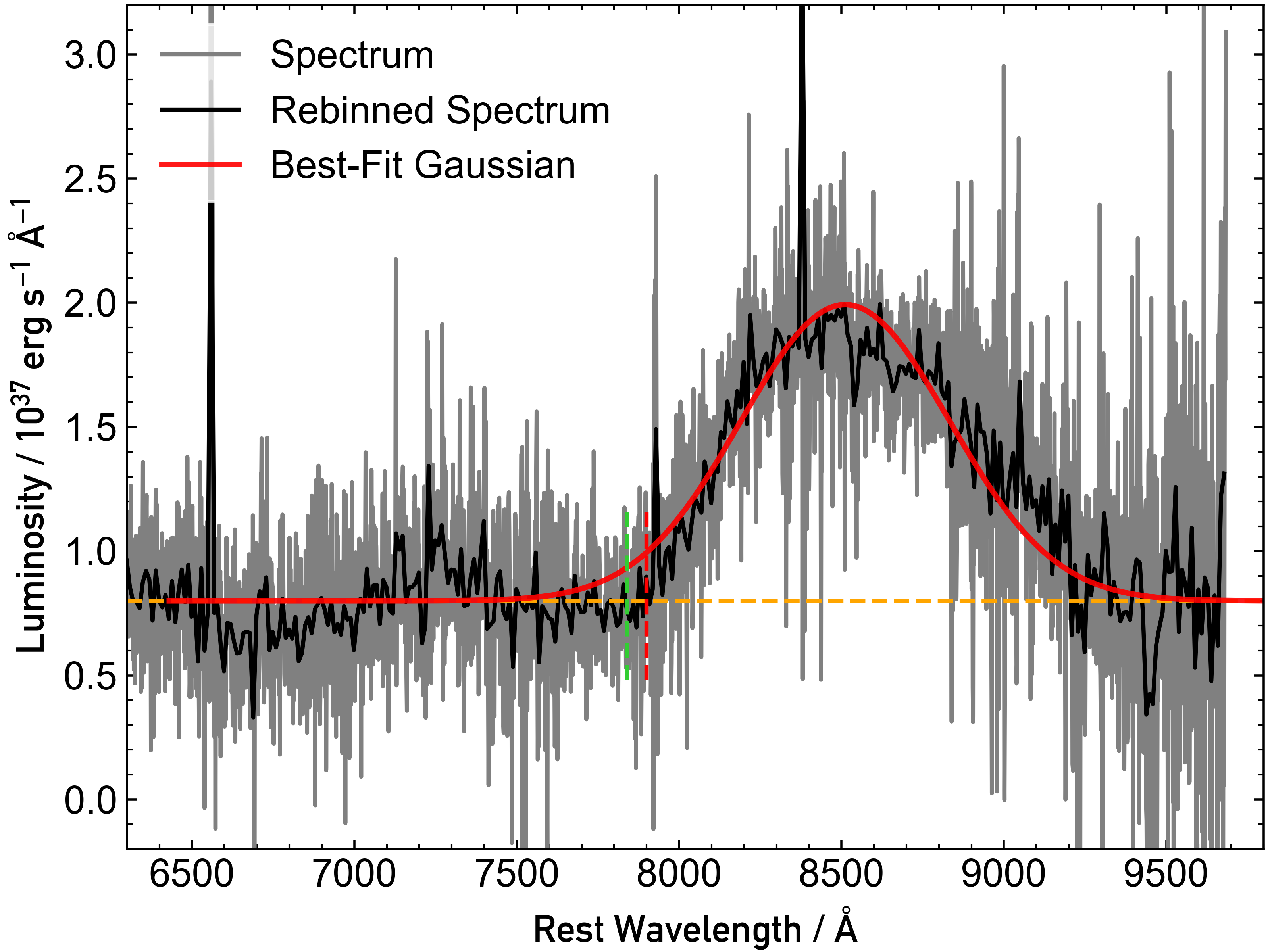}
    \caption{Comparison of our best-fit Gaussian to the observed emission feature in the +14\,d spectrum. The grey line represents the original spectrum, and the black line over-plotted represents the same spectrum but rebinned to a dispersion of 10\,\AA\,pix$^{-1}$. The red line represents our best-fit Gaussian. The horizontal orange dashed line corresponds to where we estimated the continuum of the spectrum to lie. The small vertical red dashed line at the blue end of the emission feature corresponds to where we estimate the blue wing of the feature has blended into the continuum. From this point, we can estimate the maximum velocity of the ejecta producing this feature. The vertical dashed green line represents where the blue wing blended into the continuum in the +7\,d spectrum, determined by our P-Cygni fit.}
    \label{Figures/AT2018kzr_gaussian_fits.pdf}
\end{figure}

In the +14d spectrum, we observe the \CaII\ feature to be purely in emission. We fit a Gaussian profile to the emission feature, to further estimate the velocity of the ejecta material. By assuming the feature is solely produced by the three transitions of the \CaII\ triplet, we can estimate the ejecta velocity, and the velocity offset; i.e. the velocity we observe the material to be travelling towards or away from us, relative to the bulk motion of the rest of the expanding material. We began by generating three individual Gaussians centred on the rest wavelengths of the three \CaII\ triplet lines, and assigned each of them the same width and a height which was representative of their relative strengths (if we assume the material is optically thin). We then sum these and fit the observed feature with this combined Gaussian. Figure \ref{Figures/AT2018kzr_gaussian_fits.pdf} shows the corresponding best-fit composite Gaussian obtained. 

\par
From this simple model, we estimate where the emission feature blends into the continuum. This point is on the blue wing of the emission feature at $\sim$\,7900\,\AA\ (marked by a small vertical dashed red line in Figure \ref{Figures/AT2018kzr_gaussian_fits.pdf}). From this, we obtain a maximum ejecta velocity, $\mathrm{v_{ej}}\sim$\,23700\kms, and an offset velocity (blueshift) $\mathrm{v_{offset}}\sim-$2300\kms. The offset velocity is with respect to the rest wavelength of the combined feature i.e. we have to shift each of the three components by $-$2300\kms to match the peak of the observed feature. 

\par
The maximum velocity from the emission line is similar to the maximum velocity implied by the absorption trough of the P-Cygni profile in the +7\,d spectrum. This is consistent with both features being from the same high velocity material, which has gone from a Sobolev optical depth of $\tau\sim$\,1.3, to optically thin between the two epochs. Hence, we have consistent evidence of a high-velocity component of the ejecta that is producing this observed feature, but is not contributing to the spectra at earlier epochs. Our offset velocity indicates that there is perhaps some inhomogeneity in the ejecta material, where we see excess material travelling with a component in our line of sight. This offset velocity is only a small fraction of the maximum observed ejecta velocity ($\sim$\,9.7\%) and so is likely not an issue with our interpretation of the emission feature being produced by the \CaII\ triplet.

\par
Similar blue shifted \CaII\ lines in the nebular phase of other fast fading transients have been found previously, perhaps indicating a similar physical cause. Two examples include SN2012hn \citep[][]{Valenti2014}, and SN2005E \citep[][]{Perets2010, Waldman2011}. \cite{Foley2015} studied 13 of these Ca-rich objects, and as part of their work, determined the velocity offsets for the [\CaII] $\lambda \lambda$7291, 7324 lines. They found that 3 of their sample (SN2003dg, SN2005cz and SN2012hn) had offset velocities >\,$-$1000\kms, with SN2012hn exhibiting the largest offset velocity, at $-1730$\kms. The offset velocity we obtain for our \CaII\ triplet feature ($\sim$\,$-$2300\kms) is larger than previous Ca-rich objects, by $\sim$\,33\%. An even larger offset velocity has been observed for SN2019bkc \citep[][]{Chen2020, Prentice2020}, which exhibits an offset velocity of $\sim-10000$\kms, for the Ca H\&K lines, the [\CaII] $\lambda \lambda$7291, 7324 lines, and the \CaII\ triplet.

\section{Discussion and interpretive analysis} \label{Discussion and interpretive analysis}

\subsection{Oxygen as the dominant element} \label{Oxygen as the dominant element}

\par
The dominant element, by mass fraction, in all of the models is O, which makes up $\sim$\,70\% of our ejecta material. However, the Kromer plots for all three models exhibit no significant emission or absorption features due to any ion of O. The emitting region is too cool to produce the optical \OII\ lines seen in many hot, blue superluminous supernovae \citep[these may even require non-thermal excitation;][]{Mazzali2016}, and too hot for \OI\ lines to form. This would hint that we have no need for it in the model. However, if we remove O and drop the total amount of material in the ejecta by $\sim$\,70\%, the spectrum appears significantly different. This is due to there being many fewer free electrons in the ejecta material when O is removed, as O is effectively acting as an inert electron donor. Much of the O is ionised and provides free electrons for electron scattering processes in the spectrum, without producing specific O features (which are not observed). Additionally, if everything else remains equal, a higher free electron density will lead to lower ionisation states in the model. To demonstrate that free electron donation is occurring, we calculated models with an alternative filler element i.e. one that donates free electrons, but also is not likely to produce spectral lines in the optical regime. We chose He, but could have also used C, Ne, or some mixture of the three, as these are all elements that are predicted and/or observed in significant quantities in different supernova systems.

\par
\begin{figure}
    \centering
    \includegraphics[width=\linewidth]{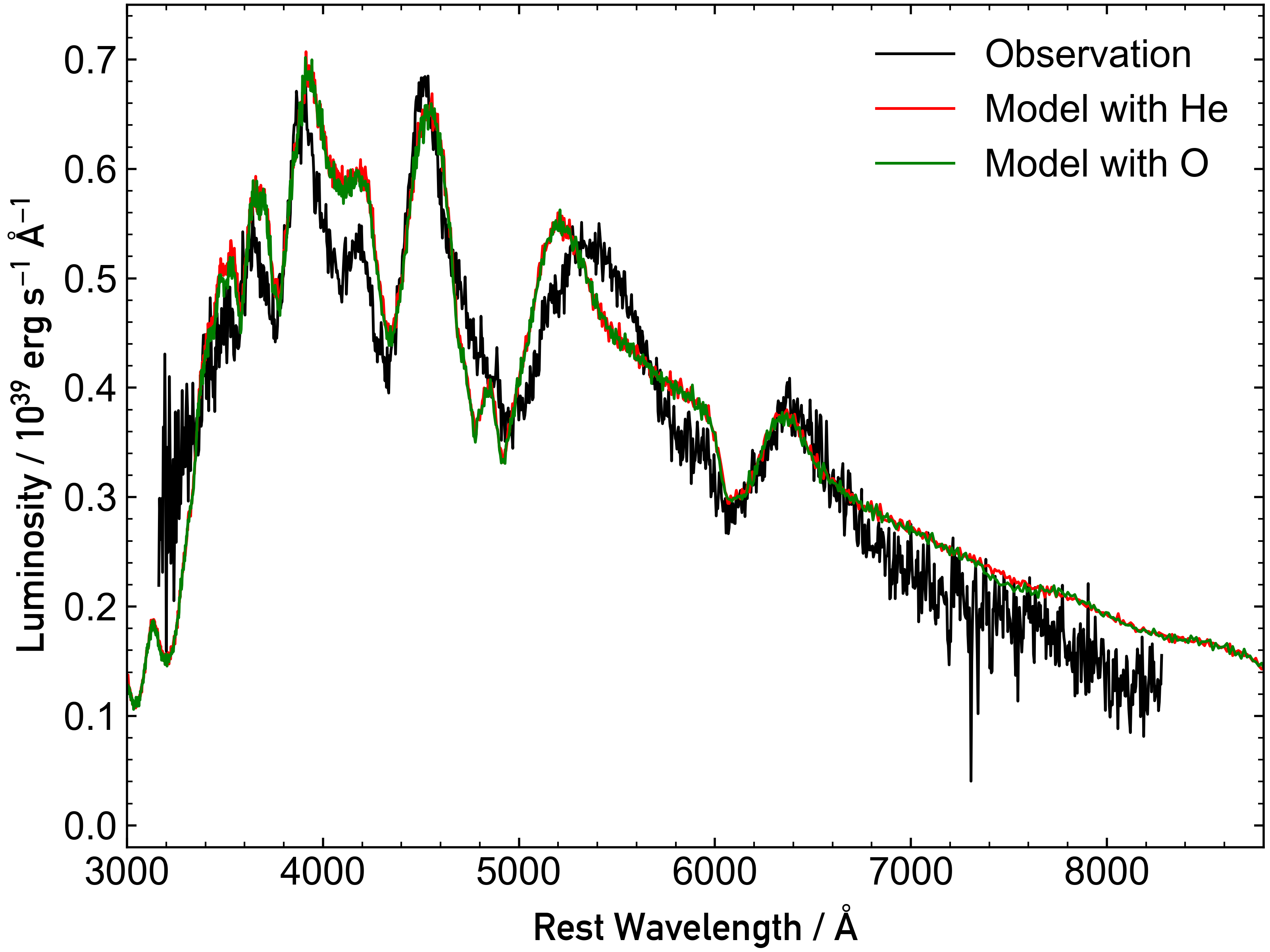}
    \caption{Comparison of the +3.8\,d spectrum with the best-fit model previously presented in Figure \ref{Figures/Epoch_3_spectral_comparison+kromer_plot.png}, and the new best-fit model, with He as the filler element instead of O. It is clearly evident that the fits almost perfectly agree with each other, and the observed spectrum.}
    \label{Figures/Epoch_3_model_with_0-92_He+7x_rho0+reduced_elements_virtual_vs_observed_spectrum_zoomed.png}
\end{figure}

We varied the total amount of material above the photosphere, and varied the ratio of the filler element relative to the other elements. We imposed the restriction of maintaining the mass fraction ratio of the elements that contribute to the spectral features. These ratios were taken from our best-fit models (see Table \ref{Table of Composition}). We were able to obtain a good model for the +3.8\,d spectrum with He as our filler element, as demonstrated in Figure \ref{Figures/Epoch_3_model_with_0-92_He+7x_rho0+reduced_elements_virtual_vs_observed_spectrum_zoomed.png}. While we can not rule out a significant amount of He being present, we propose that O is more likely and a physically motivated choice due to the significant fraction of IMEs in the ejecta that are required to model the observed features \citep[see][]{Hoflich1998, Seitenzahl2013}.

\subsection{Limits on carbon, sodium and sulphur} \label{Limits on carbon, sodium and sulphur}

\par
To further constrain the composition, we estimated upper limits on three elements (C, Na and S) that have no visible lines in any of the spectra. C and S are produced by the $\alpha$-process, while Na is synthesised via carbon burning. They are of high abundance in solar system ratios and are commonly produced in various core-collapse and thermonuclear explosion scenarios. Hence it is of interest to determine if their lack of observed lines can place an upper limit on their mass fraction in the ejecta.

\par
\begin{figure}
    \centering
    \includegraphics[width=\linewidth]{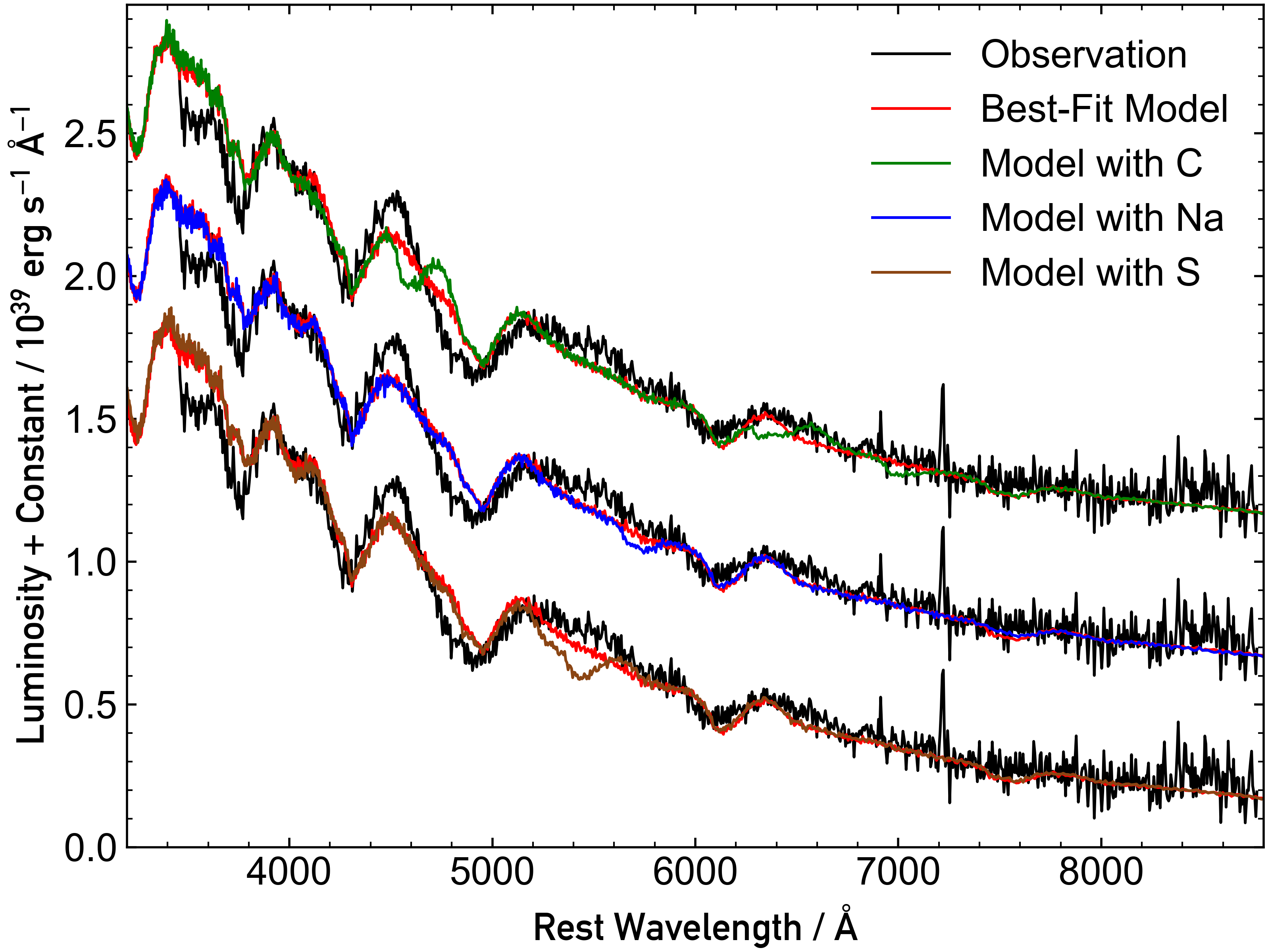}
    \caption{Comparison between the +2.8\,d spectrum, our associated best-fit model, and three models identical to our best fit, but with 8\% of the O mass fraction replaced with C (top), 40\% replaced with Na (middle), and 2\% replaced with S (bottom).}
    \label{Figures/upper_limits_for_S+C_plot.pdf}
\end{figure}

For these tests we reverted to the best-fit models that used O as the filler. We replaced various mass fractions of O with the same mass fraction of the element we were considering; i.e. to calculate a model with C mass fraction of 2\%, we removed 2\% of the total mass fraction from O, and kept all the other elements equally abundant. The bottom spectrum in Figure \ref{Figures/upper_limits_for_S+C_plot.pdf} illustrates the effect that adding in a 2\% mass fraction of S has on our best-fit model. We clearly do not observe the S feature at the model wavelength (5200--5600\,\AA) in the observed spectrum, which we attribute to the ‘w’ \SII\ feature at 5640\,\AA, and so this sets a robust constraint on the mass fraction of S that can be present. We found that we could also constrain the amount of C to <\,8\% (see the top spectrum in Figure \ref{Figures/upper_limits_for_S+C_plot.pdf}). There are three C features present in the model, at $\sim$\,4600\,\AA, $\sim$\,6400\,\AA\ and $\sim$\,7000\,\AA, due to the \CII\ $\lambda \lambda$4745, 6580 and 7234 lines. The amount of Na in the spectrum was harder to constrain as our best-fit models at all three epochs could accommodate a large mass fraction of Na without producing strong features. The middle spectrum in Figure \ref{Figures/upper_limits_for_S+C_plot.pdf} shows the effect of adding 40\% Na to the spectrum. The only noticeable effect is a weak feature at 5700\,\AA, which we attribute to the \NaI\ D $\lambda \lambda$5890, 5896 lines. Table \ref{Table of Composition} contains the upper mass limits we obtained for C and S in our three models. Na is not included as it was very unconstrained for all of our models.

\section{Discussion of progenitor systems} \label{Discussion of progenitor systems}

\par
Given the constraints on the chemical composition of the ejecta, we now discuss the four progenitor systems and explosion mechanisms that we consider to be most viable for AT2018kzr. Three of these were previously summarised by \cite{McBrien2019}; an ultra-stripped core-collapse supernova, accretion induced collapse of a WD, and a WD--NS merger (or WD--BH merger). We also include discussion on a fourth explosion mechanism; explosion of a double WD merger remnant. We begin by noting that some systematic uncertainties arise from both our spectroscopic modelling, and the lightcurve modelling performed by \cite{McBrien2019}.  $\textsc{tardis}$ has some intrinsic uncertainty for estimates of the ionisation and excitation levels for all  elements in the model. This may introduce systematic uncertainties in the determined composition. The lightcurve modelling of \cite{McBrien2019} is based upon a simple approximation for homologously expanding material, with constant opacity. The fairly simple physics employed is likely to result in a systematic uncertainty in the ejecta mass. Accordingly, we suggest that all specific values extracted from the modelling should be treated with some leeway (in the following we will adopt a factor of $\sim$\,2 as indicative of such limitations).

\par
The results from Sections \ref{Photospheric phase spectra}, \ref{Post-photospheric phase spectra} and \ref{Discussion and interpretive analysis} can be summarised as follows.  We require a significant mass fraction of Fe to fit the three photospheric spectra consistently ($3.5\pm0.4$\%), and a well-constrained (but much smaller) mass fraction of Ni and Co of $0.25\pm0.1$\% and $0.35^{+0.3}_{-0.1}$\% respectively (see Table\,\ref{Table of Composition}). The high mass fraction of Fe is unusual but appears a robust conclusion. The three early spectra are both blue and hot and simultaneously have strong absorption features due to \feii\ and \feiii. It is unusual, if not unprecedented, to find this combination since at $T_{\rm eff}\sim$\,13000\,K, these ions produce relatively weak photospheric features (for compositions with solar Fe ratios). Any explosion scenario must explain this very high Fe/(Ni+Co) ratio. It is normally assumed that most, if not all, Fe that is observed in supernovae and transients is \Fe\ from the decay chain of \Ni\ and \Co\ (or pre-existing, as in type IIP SNe). However, if  the epoch of our first model is 6.5 days after explosion, then 47.6\% of any explosively formed \Ni\ would still persist, with 50.7\% converted to \Co, and only $\sim$\,1.7\% could have decayed to stable \Fe. The mass fractions imposed by our models limit the amount of Ni and Co to $\sim$\,10\% of the mass fraction of Fe. Hence our measured composition is highly discrepant, by a factor $\sim$\,300, with the expected ratios of Ni, Co and Fe if they were synthesised from explosively produced \Ni.

\par
Furthermore, our model for the +1.9\,d spectrum requires $1.39\times10^{-2}$\,\msun\ of ejecta material  (above the opaque core) to form the synthetic spectrum, and 3.5\% of this is made up of Fe ($4.85\times10^{-4}$\,\msun). If we assume all the Fe in our first model is \Fe\ produced by radioactive \Ni\ decay, and we also trust our model explosion epoch, then we would need an initial \Ni\ mass of $2.89\times10^{-2}$\,\msun. This is larger than the total mass of material required to form the spectrum, and hence is an unphysical solution. We conclude that the spectrum must be intrinsically Fe-rich, and that this Fe must be another stable isotope, plausibly $^{54}$Fe.

\par
In Table \ref{Table Summarising Progenitor System Properties} we summarise the pros and cons of each progenitor system in the context of the observational constraints we now have for AT2018kzr, and we discuss these in detail in the next four subsections. 

\begin{table*}
\centering
\caption{Summary of the pros and cons of each of the progenitor systems considered for AT2018kzr.}
\begin{threeparttable}
\centering
\begin{tabular}{ccccc}
\hline
\hline
&{\begin{tabular}[c]{@{}c@{}}Ultra-Stripped \\ Core-Collapse SN\end{tabular}}  &{\begin{tabular}[c]{@{}c@{}}Accretion Induced \\ Collapse of an ONe WD\end{tabular}}  &{\begin{tabular}[c]{@{}c@{}}Double WD \\ Merger Systems\end{tabular}}   &{\begin{tabular}[c]{@{}c@{}}Merger of a WD \\ with an NS/BH\end{tabular}} \\
\hline
Peak Brightness       &Bad      &Possible\tnote{a}  &Good     &Possible\tnote{b}     \\
Lightcurve Evolution  &Bad      &Possible\tnote{c}  &Good     &Good                  \\
Velocity of Ejecta    &Good     &Bad                &Good     &Good\tnote{d}         \\
Ejecta Mass           &Good     &Good               &Good     &Possible\tnote{e}     \\
Bulk Composition      &Good     &Bad                &Bad      &Good                  \\
Fe/Ni Ratio           &Bad      &Bad                &Bad      &Good                  \\
\hline
\end{tabular}
\begin{tablenotes}
      \small
      \item[a] Capable of producing a magnetar, which can power the lightcurve.
      \item[b] Capable of reaching the observed luminosity of AT2018kzr, with the help of high-velocity winds launched off the disk.
      \item[c] Lightcurve evolution too fast in the models discussed, but a higher ejecta mass would act to extend the lightcurve.
      \item[d] The bulk velocity of the ejecta material agrees with our photospheric estimates for ejecta velocity.
      \item[e] The ejecta mass obtained from lightcurve modelling is lower than that estimated from the models discussed, but these values are dependent on the specific merger, which make it difficult to get good approximations.
    \end{tablenotes}\label{Table Summarising Progenitor System Properties}
\end{threeparttable}
\end{table*}

\subsection{Ultra-stripped core-collapse supernova} \label{Ultra-stripped core-collapse}

\par
\cite{Tauris2015} define an ultra-stripped SN as an exploding star that has been stripped by a degenerate companion (the amount of stripping is expected to be more than is possible by a non-degenerate companion). Typically, ultra-stripped SNe have pre-explosion envelope masses of $\lesssim$\,0.2\,\msun, but in some cases can have envelopes as low as $\sim$\,0.008\,\msun. Explosions of these objects will have correspondingly small ejecta masses, and therefore fast rise times, compared with normal core-collapse lightcurves. In particular, they are predicted to be fast-evolving, and fainter than typical type Ibc SNe. \cite{Tauris2015} have no spectroscopic predictions for ultra-stripped SNe, but the ejecta composition are expected to be broadly similar to that of type Ibc SNe, as they arise from similar progenitor systems.

\par
\cite{Moriya2017} present the nucleosynthetic yields they obtain for their ultra-stripped core-collapse model, which was based on the ultra-stripped progenitor presented in \cite{Tauris2013ultrastrip}. They evolve a star from the He zero-age main sequence (ZAMS), with the mass-loss rates expected from binary interaction (Roche lobe overflow). This results in a 1.5\,\msun\ stripped He star, with a CO core of 1.45\,\msun. This progenitor was then evolved until it collapsed, and the ejecta properties were determined for different ejecta energies and masses. The composition predicted by their models broadly agree with our composition, except for the \Ni\ production (they predict a range of \Ni\ abundances for the different models, with fractions between 0.13--0.20 of the total ejecta). Also, the lightcurves decline much too slowly to replicate the behaviour of AT2018kzr.

\par
\cite{Hachinger2012} present a model composition for SN1994I, a well observed, low-mass, stripped-envelope type Ic SN (e.g. see their Figure 7). They derive a composition for SN1994I empirically, and so we compare their results to ours, since we have adopted a similar approach. Additionally, the comparison to a Ic SN is useful as our spectra have some similarities with the spectra of type Ic SNe (see Figure \ref{Figures/AT2018kzr_vs_Ic_SNe.png}). The main difference between our method to derive the composition for AT2018kzr, and those of \cite{Hachinger2012} is that we have used a uniform, one-zone model, whereas they have a layered structure for their ejecta. Aside from this, our composition looks strikingly similar for the IMEs. In particular, their composition in the 9500--13500\kms velocity range appears almost identical to ours, with the mass dominated by O. There is $\sim$\,5\% C, and $\sim$\,few percent Na, both of which are permitted by our model. In this region, they infer significantly more Ni and less Fe than we have in our models. To summarise, if we had a type Ic SN with the composition \cite{Hachinger2012} determine for SN1994I, but with the outer layers stripped off (>\,13500\kms), and the photosphere located at 9500\kms, then our envelope composition would generally match the IMEs. 

\par
The lightcurves of type Ic (and also ultra-stripped core-collapse) SNe are thought to be powered by radioactive \Ni\ decay. The inclusion of Ni in spectrophotometric models of type Ic explosions is mainly for this reason, as well as to provide a mechanism to produce Fe, which contributes significantly to the photospheric and nebular spectra. As shown by \cite{McBrien2019}, the lightcurve of AT2018kzr is not compatible with \Ni\ powering.

\par
As discussed above, we conclude that the Fe required to form the strong absorption lines is unlikely to be \Fe\ and must be another stable isotope ($^{54}$Fe). Such ratios of Fe isotopes, and Fe/(Ni+Co) ratios are not predicted in any core-collapse mechanism and hence this makes the ultra-stripped SN scenario unlikely. The fact that \cite{McBrien2019} found that a central powering mechanism is required to produce the observed luminosity also disfavours the ultra-stripped progenitor in a binary system. Any process that severely removes mass will also serve to decrease the angular momentum, making a rapidly rotating NS remnant unlikely \citep[][]{Muller2018}.

\subsection{Accretion induced collapse of an ONe white dwarf} \label{Accretion induced collapse of an ONe white dwarf}

\par
Accretion induced collapse (AIC) may occur when an ONe WD accretes material from its companion object and grows close to the Chandrasekhar mass, causing it to reach central densities that trigger electron capture \citep[see][for a more in depth description]{Brooks2017}. At this point, it is expected that the WD collapses. When these objects collapse, they are thought to eject very little material ($\lesssim10^{-2}\msun$), as most of the material will collapse into an NS. The amount of ejected material and its properties are dependent on the properties of the WD progenitor.

\par
\cite{Darbha2010} perform lightcurve and spectroscopic analysis of AIC models. Specifically, they modelled the accretion disk that is predicted to form around the newly formed NS. For their analysis, they considered two different ejecta masses ($10^{-2}\msun$ and $3\times10^{-3}\msun$), with three different ranges of electron fraction ($\rm Y_{\rm e}$) considered for each of these masses (0.425--0.55, 0.45--0.55 and 0.5--0.55). These models have less ejected material than AT2018kzr \citep[see][where they conclude the ejecta mass to be $\sim0.10\pm0.05\msun$]{McBrien2019}. However, the findings of \cite{Darbha2010} are still useful to compare to our results.

\par
First, they predict that the outflow material will have very high velocity ($\sim$\,0.1\,$c$), which is significantly higher than the values we determine from our modelling of the early photospheric phase spectra. Second, a defining characteristic of AIC is that the disk will process material via NSE, meaning little to no IMEs will be synthesised. This contradicts our model results, which show that we only need $\sim$\,few percent IGEs, with the rest of the ejecta dominated by IMEs. The ejecta material for the models in \cite{Darbha2010} is dominated by Ni, which we do not observe in the spectra of AT2018kzr. Additionally, the evolution time-scales for the models presented by \cite{Darbha2010} are faster than that of AT2018kzr, although our larger ejecta mass may help reconcile this, since more ejected material tends to favour slower rise times. The model lightcurves are also much dimmer than that observed for AT2018kzr.

\par
The results presented by \cite{Darbha2010} broadly agree with the predictions made by \cite{Metzger2009}. In this work, \cite{Metzger2009} conclude that the ejecta material from AIC will typically have velocities, $v_{\rm ej} \sim 0.1-0.2$\,$c$, and will be predominantly made up of Ni. The multi-dimensional simulations of AIC in WDs of \cite{Dessart2006} suggest a very modest ejecta mass of $\sim$\,$10^{-3}$\,\msun, with about 25\% of that being \Ni. Such an ejecta mass and composition is not compatible with the observed lightcurve nor the composition we derive here.

\par
The work of \cite{Dessart2007} show that a magnetar may be produced in certain AIC systems, which is capable of powering the lightcurve of AT2018kzr \citep[as shown by][]{McBrien2019}. They predict that these systems would eject significantly more material ($\sim$\,0.1\,\msun), than systems which do not produce a magnetar. They also predict negligible Ni production. However, they expect the ejected material to have an electron fraction, $Y_{\rm e} \sim$\,0.1--0.2, leading them to suggest that the ejecta of an AIC with a magnetar will be dominated by $r$-process elements, which does not agree with the composition we obtained from our spectral modelling.

\subsection{Double white dwarf merger systems} \label{Double white dwarf merger systems}

\par
\cite{Brooks2017-Fast-and-Luminous} present observational predictions for fast and luminous transients that arise from the explosion of long-lived WD merger remnants. In this scenario, a He WD merges with a more massive CO or ONe WD. Most of the mass of the He WD forms an extended envelope around the massive WD, which burns for $\sim10^{5}$ years. The massive WD accretes material from this extended envelope, and as it approaches Chandrasekhar's mass, it may undergo electron capture, and collapse to an NS. The energy released by this collapse is injected into the extended envelope surrounding the NS. The models presented by \cite{Brooks2017-Fast-and-Luminous} have luminosities of $\sim 2-3 \times 10^{43}$\ergs\ 1 day after explosion. This agrees well with our observations for AT2018kzr ($L_{\rm peak} \sim 2 \times 10^{43}$\ergs). Their lightcurves are short-lived, which again agrees with our observations of AT2018kzr.

\par
There is some variation in the photospheric velocities of the models in \cite{Brooks2017-Fast-and-Luminous}, but all models have photospheric $v \sim$\,15000\kms at 2/3 days, dropping to $\sim$\,10000\kms after 5/6 days). These velocities agree well with those obtained from our spectroscopic modelling of the photospheric spectra of AT2018kzr. The models presented have ejecta masses of 0.1 and 0.2\,\msun, which is what we expect for AT2018kzr \citep[from lightcurve modelling; see][]{McBrien2019}. However, the ejecta composition does not match what we obtain from our spectroscopic modelling. \cite{Brooks2017-Fast-and-Luminous} predict an ejecta dominated by Mg and C (60/65\% Mg, 31/27\% C for envelope masses, $M_{\rm env} = 0.1/0.2 \msun$). Our models cannot accommodate this amount of C and Mg. Their models also predict much less Si (1.5/1.4\%) than we require for our models (9$^{+3}_{-2}$\%). There are no abundances quoted for IGEs, and so we cannot comment on their abundance. This progenitor system is disfavoured based on the discrepancy between the predicted composition, and the composition we have derived for AT2018kzr from our spectral modelling.

\subsection{Merger of a white dwarf with a neutron star or black hole} \label{White dwarf - neutron star merger}

\par
A WD--NS merger results from a binary system, in which a WD and an NS inspiral as the system radiates energy in the form of gravitational waves. Eventually the orbit shrinks to the point where the WD is disrupted, and a disk of WD material is formed around the NS. According to theoretical models, the temperature and pressure in the disk facilitates nuclear burning, enabling elements up to the Fe-peak to be synthesised. A WD--BH merger forms in exactly the same way, but with a stellar mass BH instead of an NS. A number of different simulations of such mergers have been performed \citep{Metzger2012,Margalit2016,Fernandez2019,Zenati2020}, with quantitative predictions for nucleosynthesis, ejecta mass, chemical composition and velocity structure. In the following, we will summarise the results and conclusions of these theoretical works, and compare our results to the predictions made.

\subsubsection{Margalit and Metzger (2016)} \label{M+M2016}

\par
The model CO WD--NS mergers of \cite{Margalit2016} show compositions dominated by unburnt C and O, which are located in the outer region of the disk, with less abundant, heavier elements (relative to C and O) located in the inner regions of the disk, as this is where they are synthesised. The heavier elements originate at smaller disk radii, where the outflow velocity is large, and so are ejected at higher velocity compared with the unburnt C and O that is expelled.  The outflow of material from wind loss is expected to contain a significant fraction of C, but we have a strict upper limit of $<$\,8\% C from our spectral models. We suggest that the lack of C in AT2018kzr can be explained by the disrupted star being an ONe WD rather than a CO WD. Our spectral synthesis requires significantly more Mg than is synthesised in these CO WD--NS mergers. All the Mg ejected in the \cite{Margalit2016} models is  synthesised in the disk. However, ONe WDs can have small quantities of Mg in their cores \citep[$\lesssim0.05\msun$,][]{Takahashi2013}, which may help to reconcile the discrepancy between our Mg abundance, and that in the CO WD--NS mergers presented by \cite{Margalit2016}. 

\par
The amount of $^{54}$Fe synthesised in the {\fontfamily{qcr}\selectfont CO\_fid} model presented by \cite{Margalit2016} makes up $\sim$\,1.1\% of the total ejecta at the simulation end time and the amount of \Ni\ synthesised is $\sim$\,6 times less ($\sim$\,0.19\%). Our $\textsc{tardis}$ model for AT2018kzr requires 14 times more Fe than Ni, which is a factor $\sim$\,2 more than in the {\fontfamily{qcr}\selectfont CO\_fid} model of \cite{Margalit2016}. The range of CO WD mergers of \cite{Margalit2016} have ratios $1.5 \lesssim$ $^{54}$Fe/$\Ni \lesssim 7.5$, with most models producing ratios at the higher end. While our inferred ratio is higher, it is still is in good qualitative agreement with the \cite{Margalit2016} calculations. It is unusual to have more $^{54}$Fe than \Ni, and so the fact that these models by \cite{Margalit2016} are in this regime is compelling.

\par
The outflow velocity of the ejecta in these models spans a wide range. The unburnt C and O can have velocities as small as $\sim$\,2000\kms, while the heavier elements possess higher velocities. The heavier elements (elements produced from oxygen and silicon-burning) originate from smaller radii in the disk, since higher temperatures are required for them to be synthesised. The mass weighted outflow velocity for the {\fontfamily{qcr}\selectfont CO\_fid} model, <$v_{\rm w}$> $\sim$\,12000\kms, which agrees well with the velocities from our spectroscopic modelling. We do not have evidence of significantly higher velocity material in the early photospheric spectra. However, a smaller amount (by mass) of high-velocity ejecta from the inner region of the disk may explain the high-velocity \CaII\ feature we observe in the two later, post-photospheric spectra (see Section \ref{Post-photospheric phase spectra}). This second component is moving at velocities between 20000--26000\kms, with a moderate asymmetry detected through the blueshift of the emission feature ($\sim$\,2300\kms).

\par
The luminosity of this model CO WD--NS transient  would reach a peak luminosity, $L_{\rm peak} \sim$\,few $10^{40}$\ergs\ if it were powered solely by \Ni\ decay. This is significantly lower than that of AT2018kzr. \cite{Margalit2016} proposed that the luminosity of the transient can be significantly enhanced by high-velocity winds, which are launched off the disk at late times. These winds collide with the bulk of the ejecta released earlier, thermalising some fraction of the wind's kinetic energy. This is quite an efficient heating process and  has the capability of boosting the luminosity to $\lesssim 2 \times10^{43}$\ergs\ (for only 10\% thermalisation efficiency). \cite{Margalit2016} also predict that the NS may be spun  up to a period, P $\sim10$\,ms. If it has a large magnetic field ($B\gtrsim10^{14}$\,G), the resultant energy deposition from magnetar spin down could also power the lightcurve \citep[see][for a semi-analytic model of magnetar powering]{McBrien2019}. The rise time for the models in \cite{Margalit2016} agree with the explosion epoch we obtain from our spectroscopic modelling. The total amount of material ejected from their CO WD--NS models is $\sim$\,0.3\,\msun, which is somewhat larger than the mass inferred from the lightcurve modelling presented by \cite{McBrien2019}. However, within the systematic uncertainties of both approaches (which are not 3D, hydrodynamic models) a difference of a factor $\lesssim$\,2 is not a significant discrepancy. 

\subsubsection{Metzger (2012)} \label{M2012}

\par
\cite{Metzger2012} presented earlier models for a wider range of merger scenarios, including CO WD--NS, He WD--NS, CO WD--BH and ONe WD--BH mergers. In each case the mass of the compact remnant was NS = 1.4\,\msun, and BH = 3\,\msun. The ejecta velocities for the heavy elements synthesised in the inner regions of the disk are predicted to be largest for the ONe WD--BH system, with mean disk wind velocities, $\bar{v}_{\rm w} \gtrsim3\times10^{4}$\kms. However, it is predicted that most of the ejected material will reside in a single shell of material at large radii, with much lower expansion velocity. The fast ejecta originating from the inner regions of the disk are then expected to collide with this slow moving ejecta, and deposit the majority of its energy in this shell, causing it all to achieve a mean velocity, $\bar{v}_{\rm ej} \sim 10^{4}$\kms. The amount of ejected material for the models range from $\sim$\,0.3--1.0\,\msun, which is higher than that permitted by the AT2018kzr lightcurve \citep{McBrien2019}. However as noted above, the lower end of that range is compatible, given the simplicity of each of the models.

\par
\cite{Metzger2012} presented two ONe WD--BH merger models and both produced significant $^{54}$Fe mass fractions of 12\% and 22\%  and \Ni\ mass fractions of 5.4\% and 1\%. The high mass fraction of $^{54}$Fe is qualitatively similar to our finding, but the \Ni\ mass predicted is significantly higher than our spectral constraints. We obtain a ratio of  $^{54}$Fe/\Ni\ $\sim$\,14 from our spectroscopic modelling, which falls between the values of 2.2 and 22 of the \cite{Metzger2012} models. If the accretion rate by the BH was higher, it would perhaps decrease the Fe and Ni ejected (as these originate from the innermost parts of the disk), bringing their mass fractions more in line with our predictions. Additionally, if the accretion rate was higher, the total amount of ejected material would also decrease. We note that significant He is predicted (15\% and 13\%) in these ONe WD--BH mergers, and we can accommodate He as a filler, or partial filler, element (see Section \ref{Oxygen as the dominant element}). The models by \cite{Metzger2012} also produce significant amounts of Si (17-18\%) and S (7-8\%) which are factors of $\sim$\,2 and $\sim$\,4 higher than inferred from the spectra of AT2018kzr.

\subsubsection{Fern\'{a}ndez et al. (2019)} \label{F2019}

\par
\cite{Fernandez2019} present more recent and detailed models for ONe WD--BH mergers, with the mass of the BH = 5\,\msun. They evolve three models for an ONe WD--BH merger; two for a non-rotating BH and the third model with a rotating BH (with a dimensionless spin of $\chi$ = 0.8), with different inner boundaries. The maximum velocity of the ejecta material for the non-rotating, large-boundary model is $\sim1.5\times10^{4}$\kms, while for the non-rotating, small-boundary model, this increases by a factor $\sim$\,2. The material ejected should be dominated by unburnt WD material, with 24\%, 34\% and 43\% composed of burning products. The rise times of these transients are predicted to be on the order $\sim$\,few days, with peak luminosity $\sim 10^{40} - 10^{41}$\ergs. This luminosity can be boosted by circumstellar interaction or late-time accretion onto the central object. For example, \cite{Fernandez2019} point out that accretion powering \citep[as in][]{Dexter2013} could boost the model luminosity, $L_{\rm} \sim 10^{43}$\ergs, within a time to peak of $t\sim3$\,d, rather similar to AT2018kzr. The predicted ejecta mass for the ONe WD--BH models is only quoted for the model with the large inner boundary ($M_{\rm ej}$ = 0.72\,\msun), which is much larger than permitted by the lightcurve model for AT2018kzr \citep[see][]{McBrien2019}. It is worth noting that the models of \cite{Fernandez2019} synthesise less Fe than the models presented by \cite{Metzger2012} and \cite{Margalit2016}, and do not predict a larger abundance of Fe than Ni. The amount of Mg synthesised in the three models is 5\%, 5\% and 3\%, which is lower than the amount we require for our models (13$^{+10}_{-4}$\%). One of their models also over-produces Si relative to our estimate for AT2018kzr. They produce 11\%, 16\% and 11\% in their three models, whereas we require 9$^{+3}_{-2}$\% for our models.  Hence the velocity, composition and plausible luminosity of an accretion powered transient all broadly agree with what we infer for AT2018kzr, with some discrepancies as described. 

\subsubsection{Zenati et al. (2020)} \label{Z2020}

\par
Not all work on modelling WD--NS mergers predicts observational properties similar to what we have observed for AT2018kzr. For example, \cite{Zenati2020} predict that the lightcurves of these mergers will be faint and red. They present models for a range of WD--NS mergers. Their models include both He and CO WDs, with masses of 0.55, 0.63 and 0.8\,\msun, and NS masses of 1.4 and 2.0\,\msun. The bolometric luminosities are expected to reach $\leq$\,$-$15, which is lower than that observed for AT2018kzr. The models presented have rise times of $\leq$\,10 days, which is broadly consistent with the rise time of AT2018kzr. The mass of unbound material at the end of the simulations range from $8 \times 10^{-3} - 0.2$\,\msun. The upper end of this range agrees well with the amount of material needed to power the lightcurve of AT2018kzr. The models by \cite{Zenati2020} do not predict that more Fe will be synthesised than Ni, which we require for our models. However, these nuclear yields were obtained from their 3D calculations, which vary significantly from previous calculations performed in 2D. The overall SED of the models in \cite{Zenati2020} do not match that of AT2018kzr, but many of the other predicted properties broadly agree.

\section{Conclusions} \label{Conclusions}

\par
In this work we have presented spectroscopic analysis of the extraordinarily fast-evolving transient, AT2018kzr. We obtained a convincing sequence of self-consistent models that replicate the early observed photospheric spectra. From these models, we extracted properties of the ejecta, such as the velocity and composition (see Tables \ref{Table of Density Profile Parameters} and \ref{Table of Composition}). The models also provided us with luminosity estimates for the transient, and an explosion epoch. The ejecta is primarily composed of the intermediate-mass elements O, Mg and Si (70\%, 13\% and 9\%, respectively) and we set a restrictive limit on the amount of C ($\leq$\,8\%). Most importantly we find that a large mass fraction of Fe is required to explain the strong absorption features in the observed spectra (3.5\%), which are predominantly formed by \feii\ and \feiii. In addition, low Ni and Co mass fractions are required (0.25\% and 0.35\%, respectively). The most common stable Fe isotope, \Fe, is formed through the decay of \Ni$\rightarrow$\Co\ (half life of 6.1 days)  and \Co$\rightarrow$\Fe\ (half life of 77.3 days). Hence, it is not possible that the Fe observed at $\sim$\,7 days post-explosion is \Fe\ that resulted from the decay of nuclei formed at the proposed explosion time. We therefore propose that it is most likely $^{54}$Fe, which significantly constrains the progenitor scenarios. 

\par
Two post-photospheric phase spectra were also analysed, and a clear \CaII\,NIR feature was identified in these spectra, taken +7 and +14 days after first detection. The \CaII\ feature first exhibited a P-Cygni profile, before transitioning to a pure emission feature by the time the second spectrum was obtained. Estimates for the velocity of the ejecta material were made from these features (see Section \ref{Post-photospheric phase spectra}), and we found that these velocities were significantly ($\sim$\,2 times) higher than the photospheric velocity estimates from our models for the early, photospheric spectra. This indicates that there is a second, high-velocity component to the ejecta.

\par
From our models, we can constrain the properties of the progenitor system that produced AT2018kzr. We analysed four plausible progenitor systems and used our results to determine which of the four, if any, best explained AT2018kzr. Based on our analysis and discussion (see Section \ref{Discussion of progenitor systems}), we identify the WD--NS or WD--BH merger as the favoured explanation. The spectral composition matches well with that hypothesised for a WD--NS or WD--BH merger scenario, if we consider the WD to be an ONe WD. In particular, a high mass fraction of synthesised $^{54}$Fe is predicted from several of these merger calculations, in agreement with our findings. The overall composition and ejecta velocities are also in good agreement. While it is clear that the \Ni\ mass in these merger models could not make a transient as bright as we observe, means to power a relatively high luminosity transient have been discussed. \cite{Margalit2016} comment that high velocity winds in the disk may produce shocks, which are capable of injecting additional energy into the wind and boosting the luminosity to levels similar to AT2018kzr. \cite{Fernandez2019} note that in a WD--BH scenario, accretion power can also provide an extra source of luminosity which could allow the transient to rise to $L_{\rm bol}\simeq10^{43}$\ergs. We note that further work is needed to better understand the specifics of these powering mechanisms. From these results, we favour the ONe WD--NS/BH merger scenario. 

\par
Theory predicts that WD--NS merger system rates could be as high as 3--15\% of the type Ia rate \citep[][]{Toonen2018}. If this is the case, then more should be detectable. The fast evolution of the transient makes it difficult to follow for an extended period, and so they need to be identified as a potential merger candidate, and observed frequently, soon after discovery. In the early phases, AT2018kzr was characterised by a hot, blue continuum, with strong Fe absorption. This unusual combination could be used as a sign that the transient is worth follow-up observations. Our models predict features in the UV at early times, and so early UV spectral observations would be good for future transients. Additionally, our spectral sequence evolves rapidly, especially in the gap between the early NTT spectra and the later Keck spectra. Higher cadence spectra (ideally every night), would be invaluable to help explain the rapid evolution into a nebular regime. The interesting \CaII\ feature we identified in our later spectra also prove the value of obtaining late-time spectra, and follow-up observations for future candidates are strongly encouraged to try for late-time spectra. Theory predicts a diversity in these objects, and so not all mergers will look and behave identically to AT2018kzr. Additional exploration of the model space, especially in terms of the ejecta nucleosynthesis, is needed to better determine the variation these mergers could exhibit, and to identify the type of system that could produce the specific properties of AT2018kzr. More of these mergers need to be discovered to better determine their diversity, but now that we have some idea of what to look out for, we hope that AT2018kzr is just the first of many interesting WD--NS/BH mergers.

\section*{Data Availability}

\par
All data and models presented in this article are available, and can be accessed from \url{https://doi.org/10.17034/89f9e0ba-7f21-4297-955d-7df3ee466d78}.

\section*{Acknowledgements}
SJS and SAS acknowledge funding from STFC Grant Ref: ST/P000312/1. This work is based (in part) on observations collected at the European Organisation for Astronomical Research in the Southern Hemisphere, Chile as part of ePESSTO, (the extended Public ESO Spectroscopic Survey for Transient Objects Survey) ESO program  199.D-0143 - ePESSTO. Some data presented herein were obtained at the W. M. Keck Observatory, which is operated as a scientific partnership among the California Institute of Technology, the University of California and the National Aeronautics and Space Administration. The Observatory was made possible by the generous financial support of the W. M. Keck Foundation. This research made use of \textsc{Tardis}, a community-developed software package for spectral synthesis in supernovae \citep{Kerzendorf2014,Kerzendorf2019}. The development of \textsc{Tardis} received support from the Google Summer of Code initiative and from ESA's Summer of Code in Space program. \textsc{Tardis} makes extensive use of Astropy and PyNE. This research also made use of the P-Cygni fitting program available publicly at \url{https://github.com/unoebauer/public-astro-tools}. We thank Owen McBrien, Brian Metzger and Rodrigo Fern\'{a}ndez for discussions and comments.

\bibliographystyle{mnras}
\bibliography{ref} % if your bibtex file is called example.bib

\appendix
\section{Host galaxy subtraction for the late time spectra} \label{Appendix - Host galaxy subtraction for the late epoch spectra}

\par
Both the +7 and +14 day spectra from Keck LRIS were severely contaminated by the flux from the host galaxy. This is not surprising given the location of AT2018kzr on top of a high surface brightness region of its galaxy, and the flux of the transient in the difference images becoming fainter than the host \citep[see the photometry in][]{McBrien2019}. The strong host contamination is clearly visible in the spectra since we observe narrow Balmer absorption features <\,4500\,\AA, which are due to the host stellar population. Their strengths can be reasonably reproduced with stellar population model \citep[e.g. STARBURST99][]{Starburst1999}, meaning that the flux bluewards of $\sim$\,4500\,\AA\ is dominated by the host light.  As described by \cite{McBrien2019}, a high-quality spectrum of the host was taken with X-shooter on the ESO VLT at +45 days, and showed no broad lines due to the transient. We convolved the X-shooter spectrum with a broad Gaussian and rebinned it to 10\,\AA\,pix$^{-1}$, and scaled its flux to match the SDSS $griz$ photometry, producing a smooth SED representing the host galaxy. This was then subtracted from the +7 and +14 day spectra.

\par
The SDSS photometry represents the whole, extended galaxy flux, and the X-shooter spectrum samples only part of it (depending on slit orientation and extraction window during reduction). Therefore we applied two scaling parameters, and allowed them to vary to produce final spectra that produced synthetic photometry that matched the observed, image subtracted, values presented by \cite{McBrien2019}. The X-shooter spectrum was scaled by a factor $s$, then subtracted from the observed spectra, which were then scaled by a factor $f$. By allowing these to vary, for the +14\,d spectrum, we produced a final spectrum with an average photometric difference, $\Delta m = -0.04 \pm 0.18$ over $griz$ between the spectrum's synthetic photometry and the observed image subtracted results. The photometric residuals for the +7\,d spectrum were even better, with <\,0.1 mag difference in each band. This is very satisfactory, and while the flux calibration should not be trusted below $\sim$\,4500\,\AA, the lines we are interested in are in the far red region.

\bsp	% typesetting comment
\label{lastpage}
\end{document}